\documentclass[11pt]{article}

\usepackage[T1]{fontenc}
\usepackage[utf8]{inputenc}
\usepackage{lmodern}
\usepackage{microtype}
\usepackage[a4paper,margin=30mm]{geometry}
\usepackage{amsmath,amssymb,amsthm,mathtools}
\usepackage{enumitem}
\usepackage{needspace}
\usepackage{xcolor}
\usepackage{tikz}
\usepackage{placeins}
\usepackage{subcaption}
\usetikzlibrary{arrows.meta,shapes.geometric,decorations.pathreplacing,patterns}
\definecolor{proofblue}{RGB}{35,83,124}
\definecolor{prooforange}{RGB}{161,88,30}
\tikzset{
  proof U/.style={circle,draw=proofblue,fill=proofblue!10,
    minimum size=6.5mm,inner sep=0pt,line width=.7pt},
  proof V/.style={rectangle,draw=prooforange,fill=prooforange!10,
    minimum size=6.5mm,inner sep=0pt,line width=.7pt},
  proof copy U/.style={circle,draw=proofblue,fill=proofblue,
    minimum size=4.5mm,inner sep=0pt},
  proof copy V/.style={rectangle,draw=prooforange,fill=white,
    minimum size=4.5mm,inner sep=0pt,line width=.9pt},
  proof common/.style={diamond,draw=black!65,fill=black!40,
    minimum size=5mm,inner sep=0pt},
  proof fiber/.style={draw=black!45,rounded corners=2pt},
  proof forward/.style={-{Stealth[length=2mm]},proofblue,line width=.8pt},
  proof reverse/.style={-{Stealth[length=2mm]},prooforange,dashed,line width=.8pt},
  proof extra/.style={-{Stealth[length=2mm]},black,densely dotted,line width=.9pt}
}
\usepackage{hyperref}

\hypersetup{
  colorlinks=true,
  linkcolor=blue!55!black,
  citecolor=blue!55!black,
  urlcolor=blue!55!black,
  pdftitle={Uniqueness, Proximity, and Scaling in Atomic Splittable Polymatroid Congestion Games},
  pdfauthor={Tobias Harks}
}
\newtheorem{theorem}{Theorem}[section]
\newtheorem{lemma}[theorem]{Lemma}
\newtheorem{corollary}[theorem]{Corollary}
\newtheorem{proposition}[theorem]{Proposition}
\theoremstyle{definition}
\newtheorem{definition}[theorem]{Definition}
\newtheorem{remark}[theorem]{Remark}
\theoremstyle{plain}
\newtheorem*{theorem*}{Theorem}

\newcommand{\R}{\mathbb{R}}

\newcommand{\Z}{\mathbb{Z}}
\newcommand{\cB}{\mathcal{B}}
\newcommand{\cI}{\mathcal{I}}
\newcommand{\cM}{\mathcal{M}}
\newcommand{\1}{\mathbf{1}}
\newcommand{\rank}{\operatorname{rk}}

\newcounter{algorithm}

\newcommand{\algorithmheading}[2]{%
  \refstepcounter{algorithm}\label{#1}%
  \textbf{Algorithm~\thealgorithm} #2.}
\DeclareRobustCommand{\RankRestore}{%
  \hyperref[alg:rank-restore]{\ensuremath{\mathsf{RankRestore}}}}
\DeclareRobustCommand{\RoundingHalving}{%
  \hyperref[alg:rounding-halving]{\textsc{Rounding-Halving}}}
\DeclareRobustCommand{\PAlg}{%
  \hyperref[prop:hkp-algorithm]{\ensuremath{\mathsf{PAlg}}}}

\title{Uniqueness, Proximity, and Scaling in\\
Atomic Splittable Polymatroid Congestion Games}
\author{Tobias Harks\\[0.5ex]
{\small University of Passau}\\
{\small \href{mailto:tobias.harks@uni-passau.de}{\texttt{tobias.harks@uni-passau.de}}}}
\date{November 19, 2026}

\begin{document}
\maketitle
\begin{abstract}
We give a polynomial-time algorithm for exact equilibrium computation in
atomic splittable polymatroid congestion games with integral ranks and
rational player-specific affine latencies with positive slopes.  The
algorithm first computes equilibria of associated integer-splittable
games, in which demands are allocated in multiples of a common packet
size.  For arbitrary polymatroid bases, no polynomial-time algorithm for
these discrete equilibria was previously known.  We provide such an
algorithm and use progressively finer packets to recover the continuous
equilibrium exactly.
Our proof uses the multiple-exchange theorem of Oki and Schwarcz (2025)
to obtain two exchange matchings that yield equilibrium proximity
under changes of packet sizes and ranks, independently of cost coefficients.
We then bound the matching deficiency caused by rank changes and construct
compatible submodular rank roundings.  Together these give a polynomial-time
packet-halving algorithm using polynomially many submodular-function
minimizations and exact rational arithmetic.
The analysis rests on a uniqueness theorem of independent interest:
for arbitrary real polymatroid ranks and continuously differentiable,
strictly increasing, convex latencies, we prove uniqueness of equilibria,
resolving a previously open problem.
\end{abstract}

\section{Introduction}
\label{sec:introduction}

In atomic splittable congestion games, players fractionally distribute fixed demands
across resources whose costs depend on the aggregate loads.  This model has been widely applied
in traffic (Haurie and Marcotte~\cite{Haurie85}), communication
networks (Orda, Rom, and Shimkin~\cite{Orda93}; Korilis et al.~\cite{KorilisL95,Korilis1997}), and freight logistics
(Cominetti, Correa, and Stier-Moses~\cite{CCS06}).  In this paper, we consider player-specific
polymatroid bases, with submodular capacity constraints on resource subsets.
Matroid strategy spaces, a special case of this structure, also arise in
queueing and facility-location models discussed by Fujishige et
al.~\cite{FGHPZ17}.

We address two basic questions for atomic splittable polymatroid congestion games: uniqueness and computation of equilibria.
Polynomial-time equilibrium algorithms for atomic splittable games are
known for restricted classes of costs and strategy spaces.  Equilibrium
computation for congestion games more generally has a long history in
algorithmic game theory and combinatorial optimization; see Ackermann et
al.~\cite{Ackermann08}.  Cominetti, Correa, and Stier-Moses \cite{CCS06} showed that, for
player-independent affine costs, equilibria minimize a convex potential and
can be approximated in polynomial time by convex programming.  For
player-specific affine costs on networks, Klimm and Warode
\cite{KlimmWarode2025} proved that equilibrium computation is PPAD-complete.
Huang \cite{Huang11} gave a combinatorial polynomial-time algorithm for
computing an exact equilibrium on certain symmetric $s$--$t$ networks with
rational player-independent affine costs.
For singleton games with rational player-specific affine costs, Harks and Timmermans
\cite{HarksT22,HarksT22Correction} obtained a polynomial-time algorithm by
exploiting a connection to integer-splittable congestion games. They proposed the following scaling approach:
\begin{enumerate}
\item Discretize demands using packets and start with a large packet size.
\item\label{enum:2} Halve the packet size.  Freeze most of the previous
      equilibrium allocation and apply the algorithm for integer-splittable games on polymatroids of Harks, Klimm, and
      Peis \cite{HarksKP18} to the remaining, shifted residual game to restore
      equilibrium at the finer scale.
\item Add the newly schedulable demand with the corresponding demand-increase
      routine, and return to Step~\ref{enum:2} until the desired scale is reached.
\end{enumerate}
Proximity bounds of equilibria with respect to different packet sizes keep the residual packet counts and the number of moves
after each demand increase polynomially bounded.
At a high level, we extend this scaling framework from singleton games
to polymatroid bases.  This extension faces three structural obstacles.  In singleton games, a packet can move from any occupied
resource to any other available resource; polymatroid constraints restrict
these moves.

\emph{First, an exchange must compare the same player at two profiles.}
We need a redistribution that is feasible at one profile and in reverse
at the other, with source and destination resources satisfying the load
comparisons that order marginal costs.  A transfer between two resources
need not satisfy all these requirements: even two spanning trees of
$K_4$ can require several edges to move together
(Remark~\ref{rem:k4-exchange}).  Oki and Schwarcz~\cite{OkiSchwarcz25}
guarantee an exchange at least as large as the load imbalance.  We
represent allocation units by labeled resource copies in an auxiliary
matroid.  Two matchings pair the exchanged copies into feasible moves
at the respective profiles; their cycles and paths turn equilibrium
inequalities into a proximity bound (Section~\ref{sec:proximity-analysis}).

\emph{Second, coarse packet sizes require compatible submodular relaxations.}
Every rank value must be a multiple of the packet size; even a rank value
of one therefore excludes packets larger than one.  We round subset
capacities upward, add a correction depending on subset size to preserve
submodularity, and cap the result at the total demand rounded down to a
packet multiple (Section~\ref{sec:rounding-scaling}).  The resulting ranks
are compatible at every scale and coincide with the original ranks at
unit scale.  Since both ranks and packet sizes change, fixed-rank
proximity does not suffice.  We bound how rank changes reduce the size
of the exchange matchings and obtain the mixed-rank proximity theorem
(Theorem~\ref{thm:mixed-rank-proximity}).

\emph{Third, a residual equilibrium must remain an equilibrium of the full game.}
After halving the packet size, we freeze most of the previous allocation
and solve a small residual game with the new ranks.  Freezing can hide
profitable full-game moves whose source contains only frozen load.
We therefore leave an unfrozen reserve, the \emph{source buffer}.
Mixed-rank proximity locates a target equilibrium above the frozen
allocation.  Subtracting the frozen load gives a reference residual
equilibrium; the residual diameter bounds its distance from the solver's
output.  A reserve covering both bounds plus one
packet ensures that every feasible full-game packet move is available
in the residual game with the same cost change.  This proves correctness
of \RankRestore{} (Lemma~\ref{lem:rank-changing-restore}).

The uniqueness of equilibria poses a related structural question.
For general (network) strategy spaces, Bhaskar, Fleischer, Hoy, and Huang
\cite{Bhaskar15} showed that equilibria can be nonunique even with
player-independent costs.  Harks and
Timmermans \cite{HarksT18} established uniqueness for certain bidirectional flow
polymatroids and gave examples of nonuniqueness under suitable embeddings for games with at
least three players whose set systems are all nonmatroidal.  Their result
left open whether uniqueness
holds for arbitrary polymatroid base strategy spaces.

\subsection{Our Results and Techniques}
Write $n$ for the number of players and $m$ for the number of resources.
We consider $m\ge2$: with only one resource, the unique feasible profile
assigns each player's entire demand to that resource and is already an
equilibrium.

We extend polynomial-time exact equilibrium computation from singleton
games to arbitrary integral polymatroid bases given by value oracles.
The algorithm handles binary-encoded demands through residual games of
polynomial size.  We also resolve the uniqueness question for arbitrary
real polymatroid bases with continuously differentiable, strictly
increasing, convex latencies.
\paragraph{Compatible rounding and exact computation.}
For the algorithmic results we assume rational player-specific affine
costs with positive slopes and integral polymatroid ranks.  A $k$-integral
strategy allocates only multiples of a common packet size $k$.  We call $k$
\emph{compatible} with a rank function if every rank value is a multiple of
$k$.

The compatible roundings of Theorem~\ref{thm:compatible-rounding}, together
with mixed-rank proximity (Theorem~\ref{thm:mixed-rank-proximity}) and the
source buffer, let \RankRestore{} pass between consecutive scales in
polynomial time.  The algorithm starts with packets larger than all rank
values and repeatedly halves their size.  After rescaling a prescribed
compatible packet size to one, stopping at unit scale gives the following
result.

\begin{theorem*}[see Theorem~\ref{thm:k-integral-computation}]
For every rational packet size $k>0$ compatible with all player-specific ranks, a
$k$-integral equilibrium can be computed in time polynomial in $n$, $m$, the
binary input length, and the binary length of $k$, even when the number of
packets per player is exponential in the input length.
\end{theorem*}

Continuing to halve the packet size brings these equilibria within a distance
of the continuous equilibrium that shrinks linearly with the packet size.
A polynomial bound on the binary length of the equilibrium coordinates then
permits exact rational reconstruction after polynomially many phases, giving
our continuous result.

\begin{theorem*}[see Theorem~\ref{thm:exact-computation-new}]
For player-specific integral polymatroid ranks given by value oracles with
binary-encoded values and player-specific rational affine costs with positive
slopes, the unique continuous Nash equilibrium can be computed exactly in time
polynomial in $n$, $m$, the binary input length, and the running time of
submodular-function minimization.
\end{theorem*}

\paragraph{Exchange properties and uniqueness.}
We also establish uniqueness of equilibria for player-specific polymatroid base polytopes,
answering an open problem of Harks and
Timmermans \cite{HarksT18}.
\begin{theorem*}[see Theorem~\ref{thm:uniqueness}]
Every atomic splittable congestion game with continuously differentiable,
strictly increasing, convex player-specific cost functions and arbitrary
player-specific polymatroid base strategy spaces has a unique Nash
equilibrium.
\end{theorem*}
Greene's classical multiple-exchange theorem~\cite{Greene74} supplies
the nonempty exchange needed for uniqueness.  Oki and
Schwarcz~\cite{OkiSchwarcz25} add the quantitative size guarantee used
for proximity.  Section~\ref{sec:exchange} transfers these properties to
polymatroids, including real ranks; Section~\ref{sec:continuous-uniqueness}
uses the resulting simultaneous direction to compare the two equilibrium
conditions.

\paragraph{Proximity of packet equilibria.}
The packet-halving method of Harks and Timmermans
\cite{HarksT22,HarksT22Correction} relies on the fact that equilibria at two
consecutive scales are close.  Establishing an analogous statement for
polymatroid bases is the central combinatorial part of our scaling analysis.

\begin{theorem*}[see Theorem~\ref{thm:mixed-proximity}]
Let $k$ and $h$ be compatible with every player-specific rank function.  Any
equilibria $x^k$ and $x^h$ of the two packet games satisfy
\[
 \max_{e\in E}|x^k_e-x^h_e|=O\bigl(nm^2(k+h)\bigr),
 \qquad
 \max_{\substack{i\in N\\e\in E}}|x^k_{i,e}-x^h_{i,e}|
 =O\bigl(nm^3(k+h)\bigr).
\]
\end{theorem*}

The two exchange matchings form directed cycles.  Equilibrium inequalities
force each cycle to contain a resource copy on which the load difference
is small.  Deleting these copies leaves paths whose lengths are bounded
by the number of resources.  Sections~\ref{sec:feedback}
and~\ref{sec:proximity} give the counting argument.  The expanded natural
matroids are proof devices; the algorithm works directly with rank
oracles.

The proximity result is useful beyond the scaling algorithm.  At one packet
size it bounds the diameter of the packet-equilibrium set
(Corollary~\ref{cor:equilibrium-lattice-diameter}); as the packet size tends
to zero it bounds the distance from a packet equilibrium to the continuous
equilibrium (Corollary~\ref{cor:continuous-proximity}).  We also allow the
polymatroid ranks themselves to change.  If two rank functions differ by at
most $\Delta$ on every set, Theorem~\ref{thm:mixed-rank-proximity} bounds the
corresponding aggregate and individual differences by
$O(nm^3(k+h)+nm^2\Delta)$ and $O(nm^5(k+h)+nm^4\Delta)$, respectively.
For the same affine costs, a limiting argument gives the continuous
sensitivity bounds $O(nm^2\Delta)$ and $O(nm^4\Delta)$ even for
arbitrary real ranks
(Corollary~\ref{cor:continuous-rank-sensitivity}).  Thus mixed-rank
proximity also quantifies how the continuous equilibrium changes when
the feasible strategy spaces are perturbed.  All these bounds are
independent of the cost coefficients.

\subsection{Further Related Work}
Orda, Rom, and Shimkin \cite{Orda93} proved uniqueness of Nash equilibria on parallel links under
convexity assumptions and obtained further sufficient conditions for general
networks.  Altman, Ba\c{s}ar, Jim\'enez, and Shimkin
\cite{Altman02competitiverouting} established uniqueness for a class of
polynomial link costs, while Richman and Shimkin \cite{Richman07}
characterized the two-terminal topologies that guarantee uniqueness under
mild convexity conditions.  Bhaskar, Fleischer, Hoy, and Huang
\cite{Bhaskar15} gave a network example with multiple equilibria under
player-independent costs.  Moving beyond path-flow strategy spaces, Harks and Timmermans
\cite{HarksT18} derived uniqueness from bidirectional polymatroid exchange
structure; our result extends this sufficiency theorem to arbitrary
polymatroid bases.
For resource cost classes closed under positive affine transformations,
Harks~\cite{harks2026} characterized universal uniqueness of equilibria in
atomic splittable congestion games.  The characterization combines a
curvature condition with strict increase of the scalar marginal costs and
includes the uniqueness results of Altman, Ba\c{s}ar, Jim\'enez, and Shimkin
\cite{Altman02competitiverouting} as a special case.

Tran-Thanh et al.~\cite{Tran11}
proved the existence of pure Nash equilibria in integer-splittable
singleton congestion games with player-independent nonnegative,
nondecreasing, convex resource costs.
Harks, Klimm, and Peis \cite{HarksKP18} developed a pseudo-polynomial
algorithm for integer-splittable polymatroid congestion games and proximity methods for
separable convex optimization over integral polymatroids.  Bhaskar and Lolakapuri \cite{BhaskarL18} obtained algorithms for approximate
equilibria for convex costs.  For polymatroid strategy spaces,
Harks and Timmermans \cite{HarksT18Convex} used packet discretization to
compute approximate equilibria under convex Lipschitz continuous costs in
pseudo-polynomial time.

Murota's discrete convex analysis \cite{Murota:2003} allows one to interpret the integral
points of polymatroid base polyhedra within the framework of
$\mathrm{M}$-convex sets and functions.  Its exchange and scaling theory
includes Murota's multiple-exchange results for valuated matroids and
$\mathrm{M}^{\natural}$-concave functions \cite{Murota18Exchange}, as well as
proximity theorems for discrete convex optimization due to Murota and Tamura
\cite{MurotaTamura04}.  These results provide optimization counterparts of the exchange and
proximity arguments used here; our bounds compare equilibria of games with
different packet sizes or strategy spaces.
In a game-theoretic direction, Takazawa \cite{Takazawa19} generalized weighted
matroid congestion games to resource costs depending on the set of users and
studied pure equilibria, sensitivity, and the associated discrete-convex
structure.

Krogmann, Lenzner, and Skopalik~\cite{KrogmannLS23} study two-stage
facility location in which clients split their demand among facilities
to minimize total waiting time.  The client stage is an atomic splittable
congestion game with a unique, efficiently computable equilibrium.
Roughgarden and Schoppmann~\cite{RoughgardenSchoppmann15} characterize
the worst-case price of anarchy of splittable congestion games with
player-independent resource costs.  Their local-smoothness framework
yields tight bounds determined by the cost class and extends these bounds
to correlated equilibria.

\subsection{Organization}
Section~\ref{sec:model} introduces polymatroid bases, the atomic splittable game, and 
variational inequality characterizations.  Section~\ref{sec:exchange} develops the required
polymatroid exchange properties, and Section~\ref{sec:continuous-uniqueness} proves
uniqueness of Nash equilibria.  Section~\ref{sec:proximity-analysis} first establishes proximity results
for the same player-specific ranks and then compares equilibria under
changes of these ranks.
Section~\ref{sec:rounding-scaling} gives the overall rounding and scaling
algorithm.  Section~\ref{sec:open-problems} discusses extensions to weighted
games and discrete convex objectives.

\section{Atomic Splittable Polymatroid Congestion Games}
\label{sec:model}

\subsection{Polymatroid bases}

Throughout, the resource set $E$ has $m\ge2$ elements.
A \emph{polymatroid rank function} is a map
$\rho:2^E\to\R_{\ge0}$ that is normalized, nondecreasing, and submodular:
\[
 \rho(\varnothing)=0,\qquad
 S\subseteq T\Longrightarrow\rho(S)\le\rho(T),\qquad
 \rho(S)+\rho(T)\ge\rho(S\cap T)+\rho(S\cup T).
\]
For $x\in\R^E$ and $T\subseteq E$, put $x(T)=\sum_{e\in T}x_e$.
The polymatroid and its base polytope are
\begin{align*}
 P(\rho)&:=\{x\in\R_{\ge0}^E:x(T)\le\rho(T)\ \forall T\subseteq E\},\\
 B(\rho)&:=\{x\in P(\rho):x(E)=\rho(E)\}.
\end{align*}
The polymatroid is \emph{integral} if every $\rho(T)$ is integral.
The base polytope is always nonempty.  Indeed, fix an ordering
$e_1,\ldots,e_m$, set $S_0:=\varnothing$ and
$S_j:=\{e_1,\ldots,e_j\}$ for every $j\in\{1,\ldots,m\}$, and define
\begin{equation}
\label{eq:greedy-base}
  x_{e_j}:=\rho(S_j)-\rho(S_{j-1})\qquad(j=1,\ldots,m).
\end{equation}
Monotonicity and telescoping give $x\ge0$ and $x(E)=\rho(E)$;
diminishing returns gives $x(T)\le\rho(T)$ for every $T\subseteq E$.
Thus \eqref{eq:greedy-base} belongs to $B(\rho)$, which is compact and
convex by its defining constraints.  This point is a vertex, called the
\emph{greedy vertex} for the chosen ordering: its coordinate at $e_j$ is
the increase in rank when $e_j$ is added to the preceding resources.
Every vertex of $B(\rho)$ arises from some ordering in this way.

\subsection{Strategies, costs, and equilibrium}

Let $N=\{1,\dots,n\}$ be the player set and let $E$ be a common set of
$m$ resources.  Player $i$ is endowed with a polymatroid rank function
$\rho_i:2^E\to\R_{\ge0}$ and chooses a vector
\[
 x_i=(x_{i,e})_{e\in E}\in B(\rho_i).
\]
Thus $x_{i,e}$ is the amount player $i$ assigns to resource $e$, and
$\rho_i(E)$ is that player's fixed total demand.  A profile is
$x=(x_i)_{i\in N}$ in the product strategy space
\[
  X:=\prod_{i\in N}B(\rho_i).
\]
The aggregate load and the load excluding player $i$ are
\[
  x_e:=\sum_{j\in N}x_{j,e},
  \qquad
  x_{-i,e}:=\sum_{j\ne i}x_{j,e}.
\]

Resource $e$ has a player-specific cost, or latency, function
\[
  c_{i,e}:\R_{\ge0}\longrightarrow\R.
\]
Throughout the continuous model we assume that every $c_{i,e}$ is
continuously differentiable, strictly increasing, and convex.
Player $i$'s cost is the sum of own load times latency over all resources,
\begin{equation}
\label{eq:player-cost}
  C_i(x)=\sum_{e\in E}x_{i,e}c_{i,e}(x_e).
\end{equation}
A profile $x\in X$ is a \emph{Nash equilibrium} if
\[
 C_i(x_i,x_{-i})\le C_i(z_i,x_{-i})
 \qquad\text{for every }i\in N\text{ and }z_i\in B(\rho_i).
\]

For fixed $x_{-i}$, the function in \eqref{eq:player-cost} is differentiable
and strictly convex in $x_i$.  Its gradient is the marginal-cost vector
\begin{equation}
\label{eq:marginal}
  \mu_{i,e}(x)
  :=\frac{\partial C_i}{\partial x_{i,e}}(x)
  =c_{i,e}(x_e)+x_{i,e}c'_{i,e}(x_e).
\end{equation}
Indeed, for $r=x_{-i,e}\ge0$, the derivative of
$\phi(q):=q\,c_{i,e}(q+r)$ is
$\phi'(q)=c_{i,e}(q+r)+q c'_{i,e}(q+r)$, which is strictly increasing:
the first term is strictly increasing and the second is nondecreasing.
Thus each summand, and hence $C_i$ in its own strategy, is strictly convex.

For a polytope $P\subseteq\R^E$ and $x\in P$, its tangent cone is
\[
 T_P(x):=\{h\in\R^E:\text{there is an }\varepsilon>0
                    \text{ with }x+\varepsilon h\in P\}.
\]
For polyhedra this is the closed cone described by the active constraint
normals.

\begin{lemma}[Best-response variational inequality]
\label{lem:best-response-vi}
A profile $x\in X$ is a Nash equilibrium if and only if, for every player
$i$,
\begin{equation}
\label{eq:best-response-vi}
  \langle\mu_i(x),z_i-x_i\rangle\ge0
  \qquad(z_i\in B(\rho_i)).
\end{equation}
Equivalently, $x$ is a Nash equilibrium if and only if, for every player
$i$,
\[
  \langle\mu_i(x),h_i\rangle\ge0
  \qquad(h_i\in T_{B(\rho_i)}(x_i)).
\]
\end{lemma}

\begin{proof}
For fixed $x_{-i}$, the objective is differentiable and convex, so its
first-order optimality condition over $B(\rho_i)$ is necessary and
sufficient.  This gives \eqref{eq:best-response-vi}.  For a tangent direction,
substitute $z_i=x_i+\varepsilon h_i$ and divide by $\varepsilon>0$.
Conversely, each feasible difference $z_i-x_i$ is tangent, giving the
equivalence of the two formulations.
\end{proof}

\section{Polymatroid Exchange Properties}
\label{sec:exchange}

To compare the variational inequalities at two profiles, we need a direction
feasible at one strategy and in the opposite orientation at the other.
We first obtain a support-restricted exchange for matroid bases and then lift
it to polymatroids.

\subsection{Generalized exchange properties}

Let $F$ be a finite set.  A \emph{matroid} is a pair
$\cM=(F,\cI)$, where $\cI\subseteq 2^F$ is nonempty, downward closed, and
satisfies augmentation: if $I,J\in\cI$ and $|I|<|J|$, then
$I+e\in\cI$ for some $e\in J\setminus I$.  Maximal independent sets are
called \emph{bases}; their family is $\cB(\cM)$ and their common cardinality is
the rank of $\cM$.  The rank function is
\[
  \rank_{\cM}(X):=\max\{|I|:I\subseteq X,\ I\in\cI\}.
\]
We write $A-X+Y:=(A\setminus X)\cup Y$, and abbreviate singleton
deletions and additions in the usual way.  Contraction and deletion are denoted
by $\cM/X$ and $\cM\backslash X$.

If $A$ is a basis and $v\notin A$, then $A+v$ contains a unique
circuit, the \emph{fundamental circuit} $C_A(v)$, and
$A-u+v$ is a basis exactly when $u\in C_A(v)\setminus\{v\}$.

We use the following basis-exchange matching, proved by Brualdi
\cite{Brualdi68} in \emph{Comments on bases in dependence structures}.

\begin{lemma}[Fundamental-exchange matching]
\label{lem:fundamental-matching}
Let $A,A'$ be bases, put $U=A\setminus A'$ and $V=A'\setminus A$, and join
$u\in U$ to $v\in V$ when $A-u+v$ is a basis.  This bipartite graph has a
perfect matching.
\end{lemma}

We use the following strengthening of the classical multiple symmetric
exchange property due to Oki and Schwarcz
\cite[Theorem~1.6]{OkiSchwarcz25}.

\begin{theorem}[Oki--Schwarcz generalized exchange theorem]
\label{thm:oki-schwarcz}
Let $A$ and $B$ be bases of a matroid with rank function $\rank$.  For arbitrary
sets $X\subseteq A\setminus B$ and $Y\subseteq B\setminus A$, there exist
sets $P\subseteq A\setminus B$ and $Q\subseteq B\setminus A$ such that
\begin{gather*}
  X\subseteq P,\qquad Y\subseteq Q,\qquad
  |P|=|Q|\le \rank(X\cup Y),\\
  A-P+Q\in\cB(\cM),
  \qquad
  B-Q+P\in\cB(\cM).
\end{gather*}
\end{theorem}

Let $\cM=(E,\cI)$ be a matroid of rank $r$, and let $A,B\in\cB(\cM)$.
Common elements can be contracted and elements outside $A\cup B$ can be
deleted.  We may therefore work with two disjoint bases satisfying
\[
  E=A\mathbin{\dot\cup}B.
\]
For an arbitrary $S\subseteq E$, write
\[
  A_1=A\cap S,\qquad A_0=A\setminus S,
  \qquad
  B_1=B\cap S,\qquad B_0=B\setminus S.
\]

Our exchange statement is the following consequence.

\Needspace{11\baselineskip}
\begin{corollary}[Support-restricted multiple strong exchange]
\label{cor:support-restricted-strong-exchange}
Let $\cM$ be a matroid of rank $r$ with disjoint bases $A,B$ satisfying
$E=A\mathbin{\dot\cup}B$, let $S\subseteq E$, and let $A_1,A_0,B_1,B_0$ be as
above.  Suppose
\[
  \delta:=|B_1|-|A_1|>0.
\]
Then there exist nonempty sets
\[
  U\subseteq A_0,
  \qquad
  V\subseteq B_1,
  \qquad
  |U|=|V|\ge \delta,
\]
such that
\[
  A-U+V\in\cB(\cM),
  \qquad
  B-V+U\in\cB(\cM).
\]
\end{corollary}

The exchange is \emph{strong} because the same pair $(U,V)$ works
simultaneously in both bases, \emph{multiple} because $U,V$ may be sets, and
\emph{support-restricted} because $U\subseteq A\setminus S$ and
$V\subseteq B\cap S$.

\begin{proof}
Prescribe the two sets on the opposite sides of $S$ from the desired exchange:
\[
  X:=A_1,
  \qquad
  Y:=B_0,
  \qquad
  F:=X\cup Y.
\]
Because $|A|=|B|=r$,
\begin{equation}
\label{eq:bad-cut-size}
  |F|
  =|A_1|+|B_0|
  =|A_1|+r-|B_1|
  =r-\delta.
\end{equation}
Apply Theorem~\ref{thm:oki-schwarcz}.  It gives $P\subseteq A$ and
$Q\subseteq B$ with
\begin{equation}
\label{eq:cover-bad}
  A_1\subseteq P,
  \qquad
  B_0\subseteq Q,
  \qquad
  |P|=|Q|
  \le \rank(F)
  \le |F|=r-\delta,
\end{equation}
and with $A-P+Q$ and $B-Q+P$ both bases.

Take the complementary exchange sets
\[
  U:=A\setminus P,
  \qquad
  V:=B\setminus Q.
\]
The containments in \eqref{eq:cover-bad} imply
\[
  U\subseteq A\setminus A_1=A_0,
  \qquad
  V\subseteq B\setminus B_0=B_1.
\]
Moreover,
\[
  |U|=|V|=r-|P|\ge\delta>0.
\]
Finally,
\begin{align*}
  A-U+V
    &=P+(B\setminus Q)=B-Q+P,\\
  B-V+U
    &=Q+(A\setminus P)=A-P+Q.
\end{align*}
Both sets are bases by construction.
\end{proof}

\begin{figure}[!htbp]
\centering
\begin{tikzpicture}[x=1cm,y=1cm,font=\small]
  \node[font=\small\bfseries] at (2.4,3.65) {$S$};
  \node[font=\small\bfseries] at (7.2,3.65) {$E\setminus S$};
  \node[anchor=east] at (-.25,2.25) {basis $A$};
  \node[anchor=east] at (-.25,.65) {basis $B$};
  \node at (2.4,3.05) {$A_1=A\cap S$};
  \node at (7.2,3.05) {$A_0=A\setminus S$};
  \node at (2.4,1.45) {$B_1=B\cap S$};
  \node at (7.2,1.45) {$B_0=B\setminus S$};
  \fill[black!13] (0,1.65) rectangle (6.4,2.85);
  \fill[proofblue!12] (6.4,1.65) rectangle (9.6,2.85);
  \fill[prooforange!13] (0,.05) rectangle (3.2,1.25);
  \fill[black!13] (3.2,.05) rectangle (9.6,1.25);
  \draw (0,1.65) rectangle (9.6,2.85);
  \draw (0,.05) rectangle (9.6,1.25);
  \draw (4.8,1.65)--(4.8,2.85);
  \draw (4.8,.05)--(4.8,1.25);
  \draw[proofblue,thick] (6.4,1.65) rectangle (9.6,2.85);
  \draw[prooforange,thick] (0,.05) rectangle (3.2,1.25);
  \node at (2.4,2.25) {$A_1\subseteq P$};
  \node at (5.6,2.25) {$P\cap A_0$};
  \node at (8,2.25) {$U=A\setminus P$};
  \node at (1.6,.65) {$V=B\setminus Q$};
  \node at (4,.65) {$Q\cap B_1$};
  \node at (7.2,.65) {$B_0\subseteq Q$};
  \node[anchor=west,font=\footnotesize] at (0,-.45)
    {Gray regions: $P$ in row $A$, $Q$ in row $B$.};
  \node at (4.8,-1.05)
    {$|P|=|Q|\le r-\delta\quad\Longrightarrow\quad |U|=|V|\ge\delta$.};
\end{tikzpicture}
\caption{The complementary exchange in
Corollary~\ref{cor:support-restricted-strong-exchange}.
The Oki--Schwarcz theorem supplies $P\supseteq A_1$ and $Q\supseteq B_0$.
Their complements $U$ and $V$ therefore lie in the required cells $A_0$
and $B_1$.  Exchanging $U,V$ produces the same two bases as exchanging
$P,Q$, in the opposite order.}
\label{fig:support-exchange}
\end{figure}
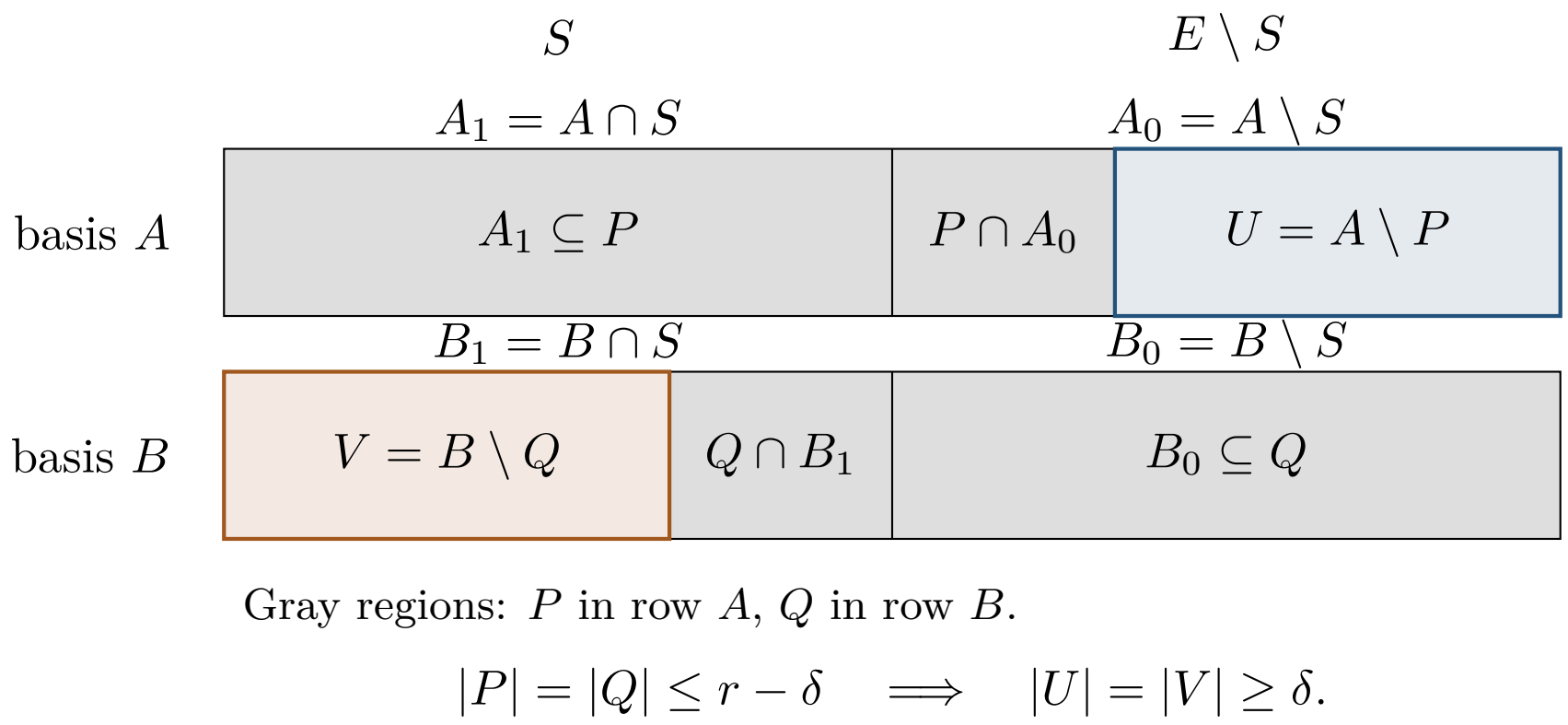

Figure~\ref{fig:support-exchange} illustrates how taking complements
places the exchanged sets on the required sides of $S$.

\begin{remark}[Qualitative exchange and uniqueness]
Greene's exchange theorem \cite{Greene74}, also stated by Oki and Schwarcz
\cite[Theorem~1.2]{OkiSchwarcz25}, already gives a nonempty simultaneous
exchange $U\subseteq A_0$, $V\subseteq B_1$: its hypothesis is
$|A_0|+|B_1|=r+\delta\ge r+1$.  This qualitative conclusion suffices for
the tangent direction and uniqueness proof below.  The exchanged sets may
contain several elements.  The quantitative bound
$|U|=|V|\ge\delta$ in
Corollary~\ref{cor:support-restricted-strong-exchange} is used in the
proximity analysis.
\end{remark}

\begin{remark}[Why several edges must move together]
\label{rem:k4-exchange}
Consider the spanning-tree matroid of the complete graph $K_4$, with each
edge labeled by its endpoints.  The sets
\[
 A=\{12,13,24\},\qquad B=\{14,23,34\}
\]
are disjoint spanning trees.  For $S=\{12,14,23\}$ we have
$A\cap S=\{12\}$ and $B\cap S=\{14,23\}$, so the imbalance is one.
A simultaneous single-edge exchange with the prescribed support would
remove $u\in\{13,24\}$ from $A$ and add $v\in\{14,23\}$, while
performing the reverse move in $B$.  All four choices fail: the pairs
$(13,14)$ and $(24,23)$ create a triangle in $A-u+v$, while
$(13,23)$ and $(24,14)$ create a triangle in $B-v+u$.
In contrast, exchanging
\[
 U=\{13,24\},\qquad V=\{14,23\}
\]
gives $A-U+V=\{12,14,23\}$ and $B-V+U=\{13,24,34\}$, both spanning
trees.  The simultaneous direction $\1_V-\1_U$ therefore has the
required support, even though no single-edge direction does.
\end{remark}

\subsection{Exchange properties for polymatroids}

\begin{definition}[Fibers of the natural-matroid construction]
\label{def:natural-matroid-fibers}
Let $\rho:2^E\to\Z_{\ge0}$ be an integral polymatroid rank function.
For each $e\in E$, choose a set $C_e$ of $\rho(\{e\})$ labeled copies of
$e$, with the sets $C_e$ pairwise disjoint.  Write
\[
 C:=\bigsqcup_{e\in E}C_e,\qquad
 \theta:C\to E,\quad\theta(c)=e\ \text{for }c\in C_e,
 \qquad C(T):=\theta^{-1}(T)=\bigcup_{e\in T}C_e.
\]
We call $C_e$ the \emph{fiber corresponding to resource $e$} and its
elements \emph{copies of $e$}.
\end{definition}

For fibers as in Definition~\ref{def:natural-matroid-fibers}, the
\emph{natural matroid} $\cM_\rho$ is the matroid on $C$ with rank function
\begin{equation}
\label{eq:natural-rank}
 r_\rho(X)=\min_{T\subseteq E}
   \bigl\{\rho(T)+|X\setminus C(T)|\bigr\},
 \qquad X\subseteq C.
\end{equation}
This is the classical natural-matroid representation of an integral
polymatroid due to Helgason \cite{helgason1974aspects}; see also Bonin, Chun, and Fife \cite{BoninCF23}.

\begin{lemma}[Projection and overlapping lifts]
\label{lem:natural-matroid}
Membership in $\cM_\rho$ depends only on the counts in each fiber.
For $X\subseteq C$, define its count projection by
\[
  \pi(X)_e:=|X\cap C_e|.
\]
For $x\in B(\rho)\cap\Z^E$, a \emph{basis lift of $x$} means a basis
$A\in\cB(\cM_\rho)$ satisfying $\pi(A)=x$.
Then
\[
 \{\pi(A):A\in\cB(\cM_\rho)\}=B(\rho)\cap\Z^E.
\]
For $x,y\in B(\rho)\cap\Z^E$, there are basis lifts
$A,B\in\cB(\cM_\rho)$ satisfying
\[
  |A\cap B\cap C_e|=\min\{x_e,y_e\}
  \qquad(e\in E).
\]
\end{lemma}

\begin{proof}
By \eqref{eq:natural-rank}, a set $I\subseteq C$ is independent exactly when
\[
 |I\cap C(T)|\le\rho(T)\qquad(T\subseteq E),
\]
which is equivalent to $\pi(I)\in P(\rho)\cap\Z^E$.  It is a basis exactly
when, in addition, $|I|=\rho(E)$, proving the projection identity.
To obtain the lifts, order each fiber and let $A$ and $B$ select its first
$x_e$ and $y_e$ copies, respectively.  There are enough copies because
$x_e,y_e\le\rho(\{e\})$.  Their count vectors are $x$ and $y$, so the
projection criterion just proved makes them bases.  The two initial segments
in each fiber overlap in exactly $\min\{x_e,y_e\}$ copies.
\end{proof}

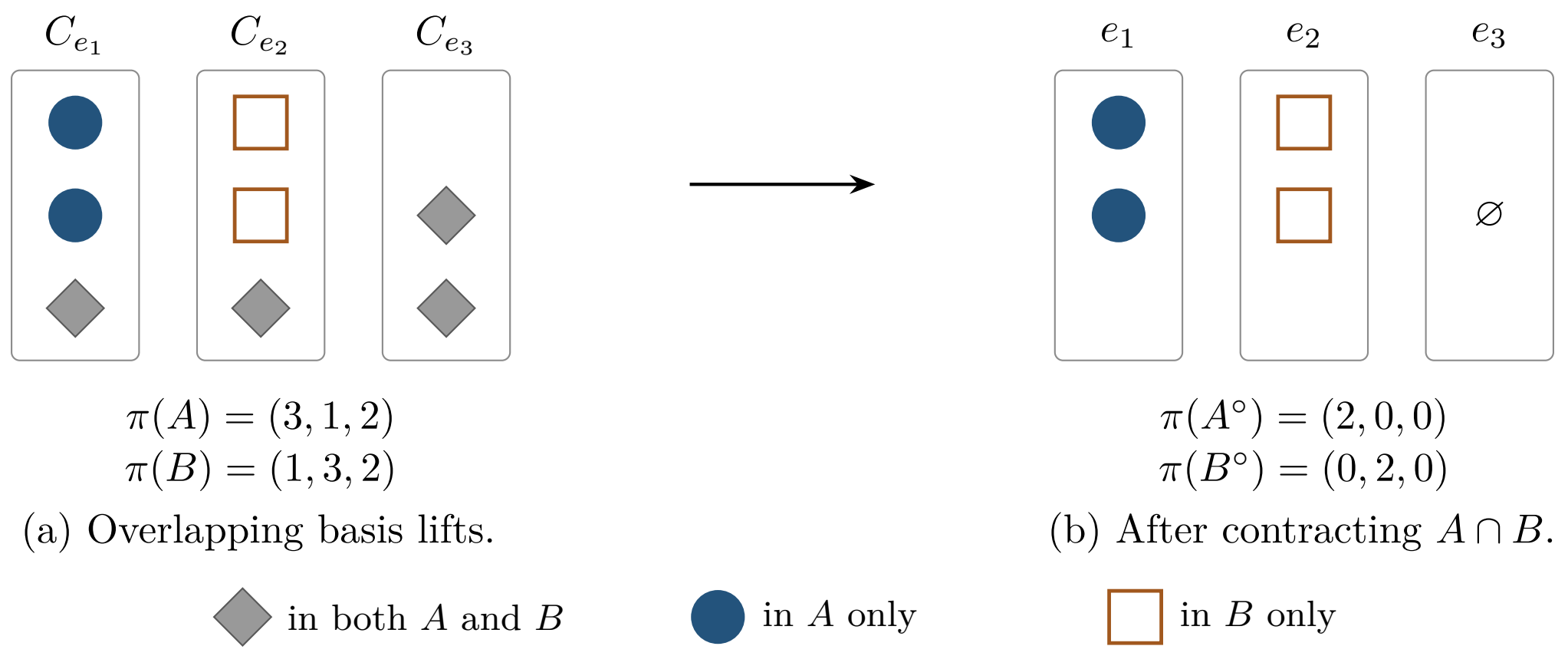
\begin{figure}[!htbp]
\centering
\begin{subfigure}[t]{.40\linewidth}
\centering\vspace{0pt}
\begin{tikzpicture}[x=1cm,y=1cm,font=\small]
  \path[use as bounding box] (0,-1.15) rectangle (4.8,3);
  \foreach \x/\e in {.8/1,2.4/2,4/3} {
    \draw[proof fiber] (\x-.55,-.05) rectangle (\x+.55,2.45);
    \node at (\x,2.75) {$C_{e_\e}$};
  }
  \node[proof common] at (.8,.4) {};
  \node[proof copy U] at (.8,1.2) {};
  \node[proof copy U] at (.8,2) {};
  \node[proof common] at (2.4,.4) {};
  \node[proof copy V] at (2.4,1.2) {};
  \node[proof copy V] at (2.4,2) {};
  \node[proof common] at (4,.4) {};
  \node[proof common] at (4,1.2) {};
  \node at (2.4,-.55) {$\pi(A)=(3,1,2)$};
  \node at (2.4,-1) {$\pi(B)=(1,3,2)$};
\end{tikzpicture}
\caption{Overlapping basis lifts.}
\label{fig:lifts-before}
\end{subfigure}\hfill
\begin{minipage}[t]{.14\linewidth}
\centering\vspace{1.45cm}
\begin{tikzpicture}
  \draw[-{Stealth[length=2.2mm]},thick] (0,0)--(1.6,0);
\end{tikzpicture}
\end{minipage}\hfill
\begin{subfigure}[t]{.40\linewidth}
\centering\vspace{0pt}
\begin{tikzpicture}[x=1cm,y=1cm,font=\small]
  \path[use as bounding box] (0,-1.15) rectangle (4.8,3);
  \foreach \x/\e in {.8/1,2.4/2,4/3} {
    \draw[proof fiber] (\x-.55,-.05) rectangle (\x+.55,2.45);
    \node at (\x,2.75) {$e_\e$};
  }
  \node[proof copy U] at (.8,1.2) {};
  \node[proof copy U] at (.8,2) {};
  \node[proof copy V] at (2.4,1.2) {};
  \node[proof copy V] at (2.4,2) {};
  \node at (4,1.2) {$\varnothing$};
  \node at (2.4,-.55) {$\pi(A^\circ)=(2,0,0)$};
  \node at (2.4,-1) {$\pi(B^\circ)=(0,2,0)$};
\end{tikzpicture}
\caption{After contracting $A\cap B$.}
\label{fig:lifts-after}
\end{subfigure}
\par\medskip
\begin{tikzpicture}[font=\footnotesize]
  \node[proof common] at (0,0) {};
  \node[anchor=west] at (.25,0) {in both $A$ and $B$};
  \node[proof copy U] at (4.1,0) {};
  \node[anchor=west] at (4.35,0) {in $A$ only};
  \node[proof copy V] at (7.7,0) {};
  \node[anchor=west] at (7.95,0) {in $B$ only};
\end{tikzpicture}
\caption{Overlapping lifts in Lemma~\ref{lem:natural-matroid}, illustrated
by two bases of the uniform matroid of rank $6$ on $8$ copies, grouped into
fibers of sizes $3,3,2$.  Contracting the four common copies leaves
$A^\circ:=A\setminus B$ and $B^\circ:=B\setminus A$, with no resource
represented in both.  In the packet application, each copy represents
$\ell$ units of load, so the remaining counts are
$(x_{i,e}-y_{i,e})_+/\ell$ and $(y_{i,e}-x_{i,e})_+/\ell$, respectively.}
\label{fig:overlapping-lifts}
\end{figure}

Figure~\ref{fig:overlapping-lifts} shows these overlaps and the disjoint
copies that remain after contracting them.

We now lift Corollary~\ref{cor:support-restricted-strong-exchange} from
matroid bases to the polymatroid base $B(\rho)$ defined in
Section~\ref{sec:model}.

\begin{lemma}[Support-restricted strong exchange direction]
\label{lem:poly-direction}
Let $x,y\in B(\rho)$ and let $S\subseteq E$ satisfy $x(S)<y(S)$.  There is a
nonzero vector $h\in\R^E$ such that
\begin{equation}
\label{eq:simultaneous-tangent}
  h\in T_{B(\rho)}(x),
  \qquad
  -h\in T_{B(\rho)}(y),
\end{equation}
and
\begin{align}
  h_e>0&\ \Longrightarrow\ e\in S,\quad x_e<y_e,
    \label{eq:positive-support}\\
  h_e<0&\ \Longrightarrow\ e\notin S,\quad x_e>y_e.
    \label{eq:negative-support}
\end{align}
If $\rho,x,y$ are rational, $h$ can be chosen rational.  If they are integral,
$h$ can be chosen integral and so that $x+h,y-h\in B(\rho)\cap\Z^E$.
\end{lemma}

\begin{proof}
First suppose $\rho,x,y$ are integral.  Form the natural matroid $\cM_\rho$
and its fibers $C_e$.  By
Lemma~\ref{lem:natural-matroid}, its bases project by fiber counts exactly onto
$B(\rho)\cap\Z^E$.

Choose basis lifts $A,B$ of $x,y$ that share exactly
$\min\{x_e,y_e\}$ copies in each fiber $C_e$.  Put $K:=A\cap B$.  In the
minor obtained by contracting $K$ and deleting the elements outside
$A\cup B$, the sets
\[
 A^\circ:=A\setminus K,
 \qquad
 B^\circ:=B\setminus K
\]
are disjoint bases, and in each fiber only one of them has remaining copies.
Define
\[
  \widetilde S:=C(S)\cap(A^\circ\cup B^\circ).
\]
The inequality $x(S)<y(S)$ becomes
\[
  |A^\circ\cap\widetilde S|<|B^\circ\cap\widetilde S|
\]
after the common copies have been contracted.  Apply
Corollary~\ref{cor:support-restricted-strong-exchange}.  We obtain
\[
  U\subseteq A^\circ\setminus\widetilde S,
  \qquad
  V\subseteq B^\circ\cap\widetilde S
\]
such that $A^\circ-U+V$ and $B^\circ-V+U$ are bases of the minor.  Reattaching
the contracted set $K$ gives bases of the original natural matroid.  Project $\1_V-\1_U$ to fiber counts and call
the resulting vector $h$.  Then
$x+h,y-h\in B(\rho)\cap\Z^E$.  A positive component can only arise from a
$B$-only copy of a resource in $S$, and a negative component can only
arise from an $A$-only copy of a resource in $E\setminus S$.  This proves
\eqref{eq:positive-support}--\eqref{eq:negative-support}.

If the data are rational, multiply $\rho,x,y$ by a common denominator, apply
the integral case, and scale the resulting direction back.

For real data, fix the combinatorial type of $(\rho,x,y)$: normalization
of $\rho$, the tight and slack monotonicity and submodularity inequalities
of $\rho$, the tight and slack base inequalities, the zero coordinates of $x,y$, the
three-way sign pattern of $y-x$, and the strict inequality $x(S)<y(S)$.
Impose all tight inequalities as equalities and keep every remaining
inequality strict, including the prescribed signs.  These are finitely many
linear conditions with rational coefficients in the coordinates of
$(\rho,x,y)$.  Their equality constraints define a rational affine space;
the strict inequalities define a relatively open subset containing the
original triple.  Rational points are dense in that affine space, so this
subset contains a rational triple
$(\widehat\rho,\widehat x,\widehat y)$ of the same type.  In particular,
$\widehat\rho$ is again a polymatroid rank.
Apply the rational case to it.  The active base-constraint normals determine
the tangent cones, giving
\[
 T_{B(\widehat\rho)}(\widehat x)=T_{B(\rho)}(x),\qquad
 T_{B(\widehat\rho)}(\widehat y)=T_{B(\rho)}(y).
\]
The preserved sign pattern therefore makes the resulting direction satisfy
\eqref{eq:simultaneous-tangent}--\eqref{eq:negative-support} for the original
data as well.
\end{proof}

\section{Uniqueness of the Continuous Equilibrium}
\label{sec:continuous-uniqueness}

\begin{theorem}[Uniqueness]
\label{thm:uniqueness}
An atomic splittable congestion game with continuously differentiable,
strictly increasing, convex player-specific costs and arbitrary
player-specific polymatroid base strategy spaces has a unique Nash
equilibrium.
\end{theorem}

\begin{proof}
Existence of equilibria follows from Rosen's theorem for concave games \cite{Rosen65}.
Suppose $x,y$ are equilibria.  The first-order variational inequalities imply,
for every
\[
  h_i\in T_{B(\rho_i)}(x_i)
       \cap\bigl(-T_{B(\rho_i)}(y_i)\bigr),
\]
that
\begin{equation}
\label{eq:vi-monotonicity}
  \langle\mu_i(x)-\mu_i(y),h_i\rangle\ge0.
\end{equation}
Indeed, feasibility of $h_i$ at $x_i$ gives
$\langle\mu_i(x),h_i\rangle\ge0$, while feasibility of $-h_i$ at $y_i$
gives $\langle\mu_i(y),h_i\rangle\le0$.

Put
\[
  d_i:=x_i-y_i,
  \qquad
  d:=\sum_i d_i.
\]
We first record the marginal-cost comparison supplied by convexity.
If $x_e<y_e$ and $x_{i,e}<y_{i,e}$, then strict increase and convexity give
\[
 c_{i,e}(x_e)<c_{i,e}(y_e),\qquad
 0\le c'_{i,e}(x_e)\le c'_{i,e}(y_e),
\]
and consequently
\begin{equation}
\label{eq:marginal-order-minus}
 \mu_{i,e}(x)<\mu_{i,e}(y).
\end{equation}
If $x_e\ge y_e$ and $x_{i,e}>y_{i,e}$, then
\begin{equation}
\label{eq:marginal-order-plus}
 \mu_{i,e}(x)>\mu_{i,e}(y).
\end{equation}
To see strictness in the second comparison, note that
$x_e\ge x_{i,e}>0$.  A differentiable, convex, strictly increasing function
has $c'_{i,e}(x_e)>0$ for $x_e>0$: otherwise monotonicity of the derivative would
make the function constant on a nontrivial initial interval.  Therefore
\[
x_{i,e}c'_{i,e}(x_e)>y_{i,e}c'_{i,e}(x_e)
                    \ge y_{i,e}c'_{i,e}(y_e).
\]

Suppose first that $d\ne0$.  We now construct the simultaneous direction
required for the contradiction.  Let
\[
  S:=\{e\in E:d_e<0\}.
\]
All profiles have the same total demand, hence $d(E)=0$, and therefore
$d(S)<0$.  Since $d(S)=\sum_i d_i(S)$, some player $i$ satisfies
$d_i(S)<0$, or equivalently $x_i(S)<y_i(S)$.  By
Lemma~\ref{lem:poly-direction}, there is a nonzero vector
\[
  h_i\in T_{B(\rho_i)}(x_i)
       \cap\bigl(-T_{B(\rho_i)}(y_i)\bigr)
\]
such that
\begin{align}
 h_{i,e}>0&\ \Longrightarrow\ d_e<0\text{ and }d_{i,e}<0,
   \label{eq:conformal-positive}\\
 h_{i,e}<0&\ \Longrightarrow\ d_e\ge0\text{ and }d_{i,e}>0.
   \label{eq:conformal-negative}
\end{align}
On a coordinate with $h_{i,e}>0$, implication
\eqref{eq:conformal-positive} and \eqref{eq:marginal-order-minus} make
$(\mu_{i,e}(x)-\mu_{i,e}(y))h_{i,e}<0$.  On a coordinate with $h_{i,e}<0$,
implication \eqref{eq:conformal-negative} and
\eqref{eq:marginal-order-plus} give the same strict inequality.  Hence every
nonzero summand in
\[
  \langle\mu_i(x)-\mu_i(y),h_i\rangle
\]
is strictly negative, contradicting \eqref{eq:vi-monotonicity}.

It remains to consider $d=0$.  If $x\ne y$, choose a player $i$ with
$d_i\ne0$ and put $S=\{e:d_{i,e}<0\}$.  Then $d_i(S)<0$.  Applying
Lemma~\ref{lem:poly-direction} to $x_i,y_i,S$ gives a nonzero simultaneous
direction with $d_{i,e}h_{i,e}<0$ on its support.  Since $d=0$, the aggregate
loads at $x$ and $y$ agree on every resource, and \eqref{eq:marginal} gives
\[
  \mu_{i,e}(x)-\mu_{i,e}(y)
  =d_{i,e}c'_{i,e}(x_e).
\]
On the support of $h_i$, the larger of $x_{i,e},y_{i,e}$ is positive, so
$x_e=y_e>0$ and $c'_{i,e}(x_e)>0$.  It follows that
$\langle\mu_i(x)-\mu_i(y),h_i\rangle<0$,
again contradicting \eqref{eq:vi-monotonicity}.  Hence $x=y$.
\end{proof}

\section{Proximity Analysis}
\label{sec:proximity-analysis}
We extend the packet-proximity analysis of Harks and Timmermans
\cite{HarksT22,HarksT22Correction} to arbitrary player-specific polymatroid
bases and to changes of the rank functions.
Sections~\ref{sec:mixed-arcs} and~\ref{sec:feedback} develop the packet
optimality conditions and the exchange-graph bounds.
We apply them first to fixed ranks in Section~\ref{sec:proximity}, pass to
the continuous limit in Section~\ref{sec:continuous-proximity}, and then
allow the ranks to change in Section~\ref{sec:mixed-rank}.

\subsection{Packet equilibria and exchange preliminaries}
\label{sec:mixed-arcs}
For the remainder of this section we impose the additional affine-cost
assumption
\begin{equation}
\label{eq:affine-specialization}
  c_{i,e}(z)=a_{i,e}z+b_{i,e},
  \qquad a_{i,e}>0,\quad b_{i,e}\in\R
  \qquad(i\in N,\ e\in E).
\end{equation}
Thus the uniqueness theorem above applies, while the exact identities
underlying the packet-proximity bounds use the stronger affine structure.
For the packet comparisons, rank functions and packet sizes are rational.
Any finite collection of these rational values has a \emph{common lattice
unit} $\ell>0$: take the reciprocal of a common denominator of their values.  Every
rank value and packet size is then an integral multiple of $\ell$.
Packet compatibility, defined below, is the stronger condition that identifies
each player's packet strategies with the integer bases of a polymatroid.
The continuous sensitivity result in Corollary~\ref{cor:continuous-rank-sensitivity}
also covers real ranks.

For any packet size $k>0$, put
\[
  B_k(\rho_i):=B(\rho_i)\cap(k\Z)^E,
  \qquad X_k:=\prod_iB_k(\rho_i).
\]
The set $B_k(\rho_i)$ consists of the lattice points of the base polytope.
Its nonemptiness is ensured by the compatibility condition below.

A packet size $k>0$ is \emph{compatible} with $\rho_i$ if
\[
  \rho_i(T)/k\in\Z
  \qquad\text{for every }T\subseteq E.
\]
Under this condition the function
\[
  (\rho_i/k)(T):=\rho_i(T)/k
\]
is an integral polymatroid rank.  Passing from resource amounts $x_i$ to
packet counts $z_i=x_i/k$ gives
\begin{equation}
\label{eq:packet-count-base}
  k^{-1}B_k(\rho_i)
  =B(\rho_i/k)\cap\Z^E,
  \qquad
  \operatorname{conv}\bigl(k^{-1}B_k(\rho_i)\bigr)=B(\rho_i/k).
\end{equation}
A \emph{$k$-integral equilibrium}, also called a packet equilibrium at
size $k$, is a profile in $X_k$ at which no
player can lower her cost by replacing $x_i$ with another point of
$B_k(\rho_i)$.  All packet games used below have compatible ranks.  Their
equilibria exist by the equilibrium-existence result for integral
polymatroid games of Harks, Klimm, and Peis
\cite{HarksKP18}.  To apply its nonnegative-cost formulation, for each
player $i$ choose
$K_i\ge\max\{0,-\min_{e\in E}b_{i,e}\}$ and replace every latency
$c_{i,e}$ by $c_{i,e}+K_i$.  This adds the constant
$K_i\rho_i(E)$ to player $i$'s objective on every feasible strategy and hence
preserves every best response and equilibrium.

For a profile $x$, the normalized costs of adding and removing one $k$-packet
on resource $e$ are
\begin{align}
  \mu_{i,e}^{+k}(x)
  &:=a_{i,e}(x_e+x_{i,e}+k)+b_{i,e},
    \label{eq:disc-plus}\\
  \mu_{i,e}^{-k}(x)
  &:=a_{i,e}(x_e+x_{i,e}-k)+b_{i,e}.
    \label{eq:disc-minus}
\end{align}
The superscripts $+k$ and $-k$ distinguish these packet marginal costs from
the continuous marginal cost $\mu_{i,e}$ in \eqref{eq:marginal}.
For $x\in X_k$, direct expansion gives the cost change of every feasible
$k$-packet move from $e$ to $f$:
\begin{equation}
\label{eq:packet-cost-difference}
 C_i\bigl(x_i+k(\1_f-\1_e),x_{-i}\bigr)-C_i(x)
 =k\bigl(\mu_{i,f}^{+k}(x)-\mu_{i,e}^{-k}(x)\bigr).
\end{equation}
For fixed opponents, each player's cost is separable and convex in her
packet counts.  Under compatibility, \eqref{eq:packet-count-base} identifies
the feasible packet counts with the integer bases of a polymatroid.
The exchange-optimality criterion of Fujishige \cite[Section~8.3]{Fujishige05}, also
stated by Harks, Klimm, and Peis \cite[Theorem~3.1]{HarksKP18}, therefore
implies that $x\in X_k$ is a $k$-integral equilibrium if and only if
\begin{equation}
\label{eq:packet-optimality}
  \mu_{i,e}^{-k}(x)
  \le
  \mu_{i,f}^{+k}(x)
\end{equation}
for every player $i$ and distinct resources $e,f$ such that
$x_i+k(\1_f-\1_e)\in B_k(\rho_i)$.

Compatibility also bounds positive exchange capacities from below.
Let $k$ be compatible with $\rho$ and let $z\in B_k(\rho)$.  For distinct
$e,f\in E$, define
\[
  \widehat c_z(e,f)
  :=\max\{\alpha\ge0:z+\alpha(\1_f-\1_e)\in B(\rho)\}.
\]
The nonnegativity and subset constraints defining $B(\rho)$ give
\begin{equation}
\label{eq:capacity-formula}
  \widehat c_z(e,f)
  =\min\left(
    z_e,
    \min_{\substack{T\subseteq E:\ f\in T,\ e\notin T}}
      \{\rho(T)-z(T)\}
  \right).
\end{equation}
Compatibility makes every term in this minimum a nonnegative multiple of
$k$, so positive exchange capacity is at least $k$.

\begin{lemma}[From copy exchanges to packet moves]
\label{lem:copy-to-packet}
Let $f/\ell$ be an integral polymatroid rank, let $k$ be compatible with $f$,
and let $A$ be a basis lift of $z/\ell$ for $z\in B_k(f)$.
For distinct resources $e,r$, take $u\in A\cap C_e$ and $v\in C_r\setminus A$.
If $A-u+v$ is a basis, then
\[
 z+k(\1_r-\1_e)\in B_k(f).
\]
In particular, a fundamental-exchange matching between $A$ and a basis
$A-U+V$, where $U\subseteq A$, $V\subseteq C\setminus A$, and no resource
is represented in both $U$ and $V$, gives a matching of feasible
$k$-packet moves at $z$.
\end{lemma}

\begin{proof}
The basis exchange projects to
$z+\ell(\1_r-\1_e)\in B(f)$ by Lemma~\ref{lem:natural-matroid}.
Its positive exchange capacity is at least $k$ by
\eqref{eq:capacity-formula} and compatibility.  The resulting move stays on the
$k$-lattice.  Applying this argument to each matched pair proves the last
claim.
\end{proof}

\subsection{Packet exchange graphs and cycle bounds}
\label{sec:feedback}

Let $x$ be a $k$-integral equilibrium on $(B(f_i))_i$ and $y$ an
$h$-integral equilibrium on $(B(g_i))_i$, with $f_i$ compatible with $k$
and $g_i$ compatible with $h$.  Both games use the same affine costs
\eqref{eq:affine-specialization}.  Throughout the proximity proofs, write
\[
 d_{i,e}:=x_{i,e}-y_{i,e},\qquad
 d_e:=\sum_i d_{i,e},\qquad s_{i,e}:=d_e+d_{i,e},\qquad t:=k+h.
\]
For a fixed player $i$, let $U,V$ be disjoint sets of resource copies, with
no resource represented in both sets.  Recall that $C_e$ denotes the copies
of resource $e$, and $C(T)=\bigcup_{e\in T}C_e$ denotes all copies of
resources in $T\subseteq E$.  Define
\[
 U_e:=U\cap C_e,\qquad V_e:=V\cap C_e.
\]
The \emph{packet exchange graph} $\mathcal D_i(U,V)$ has vertex set
$U\cup V$ and directed edges defined, for $u\in U_e$ and $v\in V_r$, by
\begin{align*}
 u\to v&\quad\Longleftrightarrow\quad
 x_i+k(\1_r-\1_e)\in B(f_i),\\
 v\to u&\quad\Longleftrightarrow\quad
 y_i+h(\1_e-\1_r)\in B(g_i).
\end{align*}
Vertices in the same fiber have identical outgoing neighbors because edge
feasibility depends only on resources.  The copies retain the multiplicities
of an exchange, while the edges record feasible packet moves at the two
equilibria.

To see how matchings enter this graph, suppose $A$ is a basis lift of
$x_i/\ell$, with $U\subseteq A$, $V\cap A=\varnothing$, and $A-U+V$ another
basis.  Apply
Lemma~\ref{lem:fundamental-matching} to these two bases: it pairs every
$u\in U$ with a distinct $v\in V$ for which $A-u+v$ is a basis.
Lemma~\ref{lem:copy-to-packet} turns each such exchange into a feasible
$k$-packet move, so these edges form a perfect forward matching in
$\mathcal D_i(U,V)$.  If $B$ is a basis lift of $y_i/\ell$ in the natural
matroid of $g_i/\ell$, with $V\subseteq B$, $U\cap B=\varnothing$, and
$B-V+U$ also a basis, Lemmas~\ref{lem:fundamental-matching} and~\ref{lem:copy-to-packet},
now applied at $B$ with packet size $h$, give a perfect reverse matching.

Call $u\in U_e$ \emph{exceptional} if $s_{i,e}\le t$, and call $v\in V_r$ exceptional if
$s_{i,r}>-t$.  Define the set of exceptional copies by
\[
 Z:=\bigcup_{e:\,s_{i,e}\le t}U_e
     \;\cup\!\bigcup_{r:\,s_{i,r}>-t}V_r
 \subseteq U\cup V.
\]
The next three lemmas explain why every cycle must meet $Z$, bound the
number of such copies, and turn those facts into a bound on $|U|=|V|$.

\begin{lemma}[Every cycle contains an exceptional copy]
\label{lem:bad-cycle}
Let $x$ be a $k$-integral equilibrium on $(B(f_i))_i$ and $y$ an
$h$-integral equilibrium on $(B(g_i))_i$, where $f_i$ is compatible with $k$
and $g_i$ with $h$.  Suppose both games have the same affine costs
\eqref{eq:affine-specialization}.  For $U,V$ as above, every directed cycle
of $\mathcal D_i(U,V)$ meets the set $Z$ of exceptional copies.
\end{lemma}

\begin{proof}
Suppose, for a contradiction, that a directed cycle avoids $Z$.  Since the
graph is bipartite, the cycle alternates between $U$ and $V$.  Consider
consecutive edges $u\to v\to u'$, with resources $e,r,e'$, respectively.
The absence of exceptional copies means
\[
 s_{i,e}>t,\qquad s_{i,r}\le-t.
\]
Subtracting the affine marginal formulas
\eqref{eq:disc-plus}--\eqref{eq:disc-minus} gives the following comparisons,
using the definition of $s_{i,e}$ and the positive slopes:
\begin{align}
 \mu_{i,e}^{-k}(x)-\mu_{i,e}^{+h}(y)
 &=a_{i,e}(s_{i,e}-t)>0,\label{eq:cross-positive}\\
 \mu_{i,r}^{-h}(y)-\mu_{i,r}^{+k}(x)
 &=a_{i,r}(-s_{i,r}-t)\ge0.\label{eq:cross-negative}
\end{align}
On the other hand, $u\to v$ is a feasible $k$-packet move at $x$, and
$v\to u'$ is a feasible $h$-packet move at $y$.  Since both profiles are
equilibria, \eqref{eq:packet-optimality} yields
\[
 \mu_{i,e}^{-k}(x)\le\mu_{i,r}^{+k}(x),
 \qquad
 \mu_{i,r}^{-h}(y)\le\mu_{i,e'}^{+h}(y).
\]
Combining the four comparisons gives the chain, with each step annotated
by its source:
\[
 \mu_{i,e}^{+h}(y)
 \underset{u\notin Z}{<}\mu_{i,e}^{-k}(x)
 \underset{\text{equilibrium at }x}{\le}\mu_{i,r}^{+k}(x)
 \underset{v\notin Z}{\le}\mu_{i,r}^{-h}(y)
 \underset{\text{equilibrium at }y}{\le}\mu_{i,e'}^{+h}(y).
\]
The value $\mu_{i,e}^{+h}(y)$ strictly increases whenever we follow two
successive edges from a $U$-vertex to the next $U$-vertex.  Traversing the
entire cycle returns to the starting vertex and would make this value
strictly larger than itself, a contradiction.
\end{proof}

For the fixed-rank comparison in Section~\ref{sec:proximity}, we take
$\widehat y_i=y_i$ and $\eta=0$.  The general form below also covers
Section~\ref{sec:mixed-rank}, where $\widehat y_i$ approximates $y_i$ in
the other base polytope $B(f_i)$.

\begin{lemma}[Counting exceptional copies]
\label{lem:bad-count}
Use the notation $d_{i,e},d_e,t,U_e,V_e,Z$ defined above.
Let $\eta\ge0$ and let $\widehat y_i\in\R^E$ satisfy
$\|\widehat y_i-y_i\|_\infty\le\eta$.  Suppose that, for every resource $e$,
\[
 \ell|U_e|\le(x_{i,e}-\widehat y_{i,e})_+,
 \qquad
 \ell|V_e|\le(\widehat y_{i,e}-x_{i,e})_+,
\]
where $(a)_+:=\max\{a,0\}$.
\begin{enumerate}[label=\textnormal{(\roman*)},leftmargin=2.6em]
 \item Let $P:=\{e:d_e>0\}$.  If every copy in $U$ represents a resource
       in $P$, and every copy in $V$ represents a resource in $E\setminus P$,
       then $|Z|\le m(t+\eta)/\ell$.
 \item If $|d_r|\le G$ for all $r$, every copy in $U$ represents the same
       resource $e$, and $d_{i,e}>G+t$, then
       $|Z|\le(m-1)(G+t+\eta)/\ell$.
\end{enumerate}
\end{lemma}

\begin{proof}
Whether a copy is exceptional depends only on its resource.  The assumed error bound
$\|\widehat y_i-y_i\|_\infty\le\eta$ implies
\[
 x_{i,e}-\widehat y_{i,e}\le d_{i,e}+\eta,
 \qquad
 \widehat y_{i,e}-x_{i,e}\le-d_{i,e}+\eta.
\]
If $U_e$ contains an exceptional copy, then
$d_{i,e}\le t-d_e$.  The size of $U_e$ is consequently bounded by
\[
 \ell|U_e|\le(x_{i,e}-\widehat y_{i,e})_+
 \le(d_{i,e}+\eta)_+\le(t-d_e+\eta)_+.
\]
Likewise, if $V_r$ contains an exceptional copy, then $-d_{i,r}<t+d_r$, and hence
\[
 \ell|V_r|\le(\widehat y_{i,r}-x_{i,r})_+
 \le(-d_{i,r}+\eta)_+\le(t+d_r+\eta)_+.
\]

In case (i), every $U$-fiber has $d_e>0$ and every $V$-fiber has $d_r\le0$.
Each exceptional fiber therefore contributes at most $(t+\eta)/\ell$ copies.
No resource occurs in both $U$ and $V$, so there are at most $m$ such fibers
altogether.  Summing gives $|Z|\le m(t+\eta)/\ell$.

In case (ii), the single resource supporting $U$ satisfies
\[
 s_{i,e}=d_e+d_{i,e}\ge-G+d_{i,e}>t.
\]
Every copy in $U$ lies outside $Z$.  The fibers supporting $V$ use only the
other $m-1$ resources, and $d_r\le G$ bounds each exceptional one by
$(G+t+\eta)/\ell$.  Summing these contributions proves (ii).
\end{proof}

The following counting lemma bounds the exchange size from the number of
exceptional copies and the matching deficiencies.

\begin{lemma}[Cycle-hitting bound with matching deficiencies]
\label{lem:partial-feedback-path}
Let $U$ and $V$ be disjoint sets of vertices, with $|U|=|V|=L$, and let
$D=(U\cup V,\mathcal A)$ be a directed graph whose edge set satisfies
\[
 \mathcal A\subseteq(U\times V)\cup(V\times U).
\]
Partition the vertices into at most $m$ fibers, each contained in $U$ or
in $V$, and assume that vertices in the same fiber have identical outgoing
neighbors.

Suppose $D$ contains a matching $M^+$ of edges from $U$ to $V$ and a
matching $M^-$ of edges from $V$ to $U$.  Assume
\[
 |M^+|\ge L-\beta^+,\qquad |M^-|\ge L-\beta^-,
 \qquad \beta^+,\beta^-\ge0.
\]
The numbers $\beta^+,\beta^-$ therefore bound the respective
\emph{deficiencies}, namely the numbers of unmatched vertices in each of
$U$ and $V$.

Let $Z\subseteq U\cup V$ be a set of vertices such that every directed cycle
of $D$ contains at least one vertex of $Z$.  Then
\begin{equation}
 2L-|Z|\le m(\beta^++\beta^-+|Z|),
 \qquad
 L\le\frac m2(\beta^++\beta^-)+\frac{m+1}{2}|Z|.
\label{eq:partial-L-bound}
\end{equation}
In particular, two perfect matchings give $L\le(m+1)|Z|/2$.
\end{lemma}

\begin{proof}
Let $H$ have vertex set $U\cup V$ and edge set $M^+\cup M^-$.  A vertex of
$U$ has at most one outgoing edge from $M^+$ and at most one incoming
edge from $M^-$.  The roles are reversed at vertices of $V$.  Thus each vertex
has at most one incoming and one outgoing edge, and $H$ has at least
$2L-\beta^+-\beta^-$ edges.

Delete the vertices in $Z$ and all their incident edges to obtain $H_0$.
Each deleted vertex is incident with at most two edges, so $H_0$ has
$2L-|Z|$ vertices and at least $2L-\beta^+-\beta^--2|Z|$ edges.  It has no
directed cycle, because such a cycle would also be a cycle of $D$ containing
no vertex of $Z$.  Since each remaining vertex still has at most one incoming
and one outgoing edge, every component of $H_0$ is a directed path, with
isolated vertices counted as paths.  A path has one more vertex than edges.
The number $p$ of these paths satisfies
\[
 p\le(2L-|Z|)-(2L-\beta^+-\beta^--2|Z|)
   =\beta^++\beta^-+|Z|.
\]

We next show that each path uses a fiber at most once.  Write a path as
$w_1\to\cdots\to w_q$ and suppose $w_a,w_b$ lie in the same fiber,
with $a<b$.  Both belong to $U$ or both belong to $V$.  Since every edge
switches between these vertex sets, $b\ge a+2$.  The edge
$w_a\to w_{a+1}$ exists, so the assumption on outgoing neighbors also
gives the edge $w_b\to w_{a+1}$ in $D$.  Together with the path segment
from $w_{a+1}$ to $w_b$, this edge forms a directed cycle with no vertex
in $Z$.  The closing edge belongs to $D$, so this cycle contradicts the
defining property of $Z$.

Each path therefore contains at most $m$ vertices.  Counting all vertices of
$H_0$ gives
\[
 2L-|Z|\le mp\le m(\beta^++\beta^-+|Z|).
\]
Rearranging proves the second inequality in \eqref{eq:partial-L-bound}.
For two perfect matchings, set $\beta^+=\beta^-=0$.
\end{proof}

\begin{figure}[!ht]
\centering
\begin{subfigure}[t]{.48\linewidth}
\centering\vspace{0pt}
\begin{tikzpicture}[x=1cm,y=1cm,font=\small]
  \path[use as bounding box] (-.05,0) rectangle (5.95,2.65);
  \foreach \shift/\tag in {0/a,3.4/b} {
    \begin{scope}[xshift=\shift cm]
      \node[proof U,double,double distance=1pt] (\tag e) at (.35,2.2) {$e$};
      \node[proof V] (\tag r) at (2.15,2.2) {$r$};
      \node[proof U] (\tag f) at (2.15,.45) {$f$};
      \node[proof V] (\tag s) at (.35,.45) {$s$};
      \draw[proof forward] (\tag e)--(\tag r);
      \draw[proof reverse] (\tag r)--(\tag f);
      \draw[proof forward] (\tag f)--(\tag s);
      \draw[proof reverse] (\tag s)--(\tag e);
    \end{scope}
  }
\end{tikzpicture}
\caption{The two matchings form directed cycles.}
\label{fig:exchange-cycles}
\end{subfigure}\hfill
\begin{subfigure}[t]{.48\linewidth}
\centering\vspace{0pt}
\begin{tikzpicture}[x=1cm,y=1cm,font=\small]
  \path[use as bounding box] (-.05,0) rectangle (5.95,2.65);
  \foreach \y/\tag in {2.2/c,.45/d} {
    \node[proof V] (\tag r) at (.6,\y) {$r$};
    \node[proof U] (\tag f) at (2.95,\y) {$f$};
    \node[proof V] (\tag s) at (5.3,\y) {$s$};
    \draw[proof reverse] (\tag r)--(\tag f);
    \draw[proof forward] (\tag f)--(\tag s);
  }
\end{tikzpicture}
\caption{Deleting $Z$ leaves directed paths.}
\label{fig:exchange-deletion}
\end{subfigure}
\par\bigskip
\begin{subfigure}[t]{\linewidth}
\centering
\begin{tikzpicture}[x=1cm,y=1cm,font=\small]
  \node[proof U,label=above:{$w_a$}] (wa) at (.65,0) {$\alpha$};
  \node[proof V,label=above:{$w_{a+1}$}] (wan) at (3.35,0) {$\beta$};
  \node[proof U,label=above:{$w_{a+2}$}] (wam) at (6.05,0) {$\gamma$};
  \node[proof V,label=above:{$w_{b-1}$}] (wbp) at (8.75,0) {$\delta$};
  \node[proof U,label=above:{$w_b$}] (wb) at (11.45,0) {$\alpha$};
  \draw[proof forward] (wa)--(wan);
  \draw[proof reverse] (wan)--(wam);
  \draw[proof forward] (wam)--(wbp);
  \draw[proof reverse] (wbp)--(wb);
  \draw[proof extra] (wb.south) to[out=-100,in=-80]
    node[below,font=\footnotesize] {closing edge $w_b\to w_{a+1}$ in $D$}
    (wan.south);
\end{tikzpicture}
\caption{Repeating a resource closes a cycle.}
\label{fig:exchange-repeat}
\end{subfigure}
\par\medskip
\begin{tikzpicture}[font=\footnotesize]
  \draw[proof forward] (0,0)--(.8,0);
  \node[anchor=west] at (.88,0) {$M^+:U\to V$};
  \draw[proof reverse] (3.75,0)--(4.55,0);
  \node[anchor=west] at (4.63,0) {$M^-:V\to U$};
  \node[proof U,double,double distance=1pt,minimum size=4mm] at (8,0) {};
  \node[anchor=west] at (8.35,0) {vertex in $Z$};
\end{tikzpicture}
\caption{The cycle and path argument of
Lemmas~\ref{lem:bad-cycle} and~\ref{lem:partial-feedback-path}.
Circles are vertices in $U$, squares are vertices in $V$, and letters identify
resources.  In (\subref{fig:exchange-cycles}), the two copies of $e$ form
$Z$.  Lemma~\ref{lem:bad-cycle} ensures that every directed cycle meets
$Z$, and its deletion gives the paths in (\subref{fig:exchange-deletion}).
In (\subref{fig:exchange-repeat}), suppose a path avoiding $Z$ contains two
copies of $\alpha$.  Their identical outgoing neighbors supply the dotted
edge, closing a cycle that avoids $Z$.  This contradiction bounds each
remaining path by $m$ vertices.}
\label{fig:exchange-paths}
\end{figure}
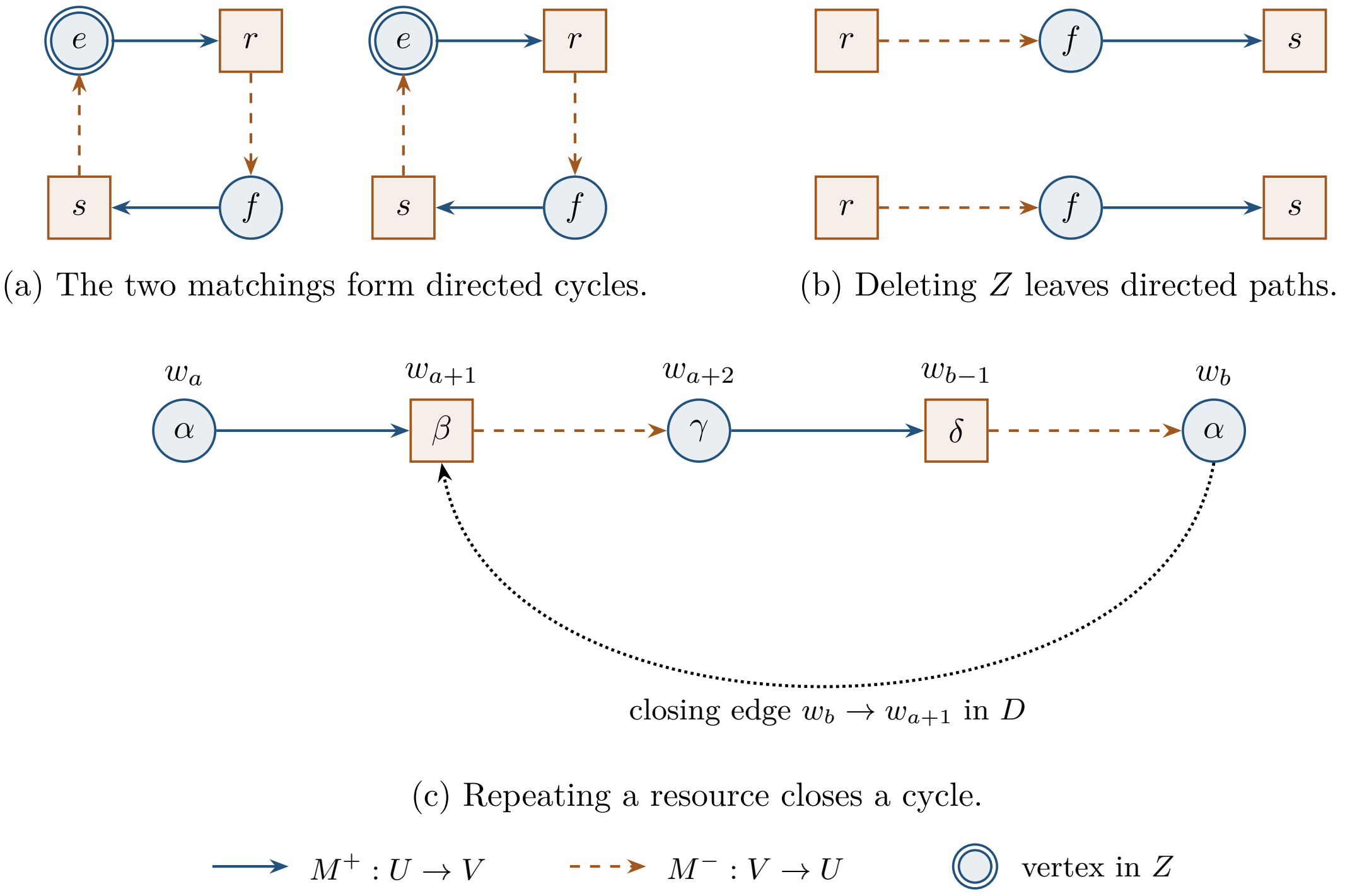

Figure~\ref{fig:exchange-paths} illustrates the deletion of $Z$ and the
cycle that would arise if a remaining path repeated a resource.

In the proximity proofs below, we apply this lemma with
$D=\mathcal D_i(U,V)$ and with $Z$ equal to the set of exceptional copies defined
above.  Lemma~\ref{lem:bad-cycle} verifies that every directed cycle
contains a vertex of $Z$, and Lemma~\ref{lem:bad-count} bounds its size.

\FloatBarrier

\subsection{Proximity for fixed compatible ranks}
\label{sec:proximity}

We first compare equilibria $x^k,x^h$ for the same ranks $\rho_i$, with
both packet sizes compatible.  Use the notation of
Section~\ref{sec:feedback} with $x=x^k$ and $y=x^h$.

\begin{theorem}[Compatible mixed-packet proximity]
\label{thm:mixed-proximity}
Let $k,h>0$ be packet sizes compatible with every player-specific rank
function.  For every pair of equilibria $x^k,x^h$,
\begin{equation}
\label{eq:aggregate-proximity}
  \max_{e\in E}|x_e^k-x_e^h|
  \le
  \frac{nm(m+1)}{2}(k+h)
  \le nm^2(k+h).
\end{equation}
Moreover,
\begin{equation}
\label{eq:individual-proximity}
  \max_{\substack{i\in N\\e\in E}}|x_{i,e}^k-x_{i,e}^h|
  \le
  (m-1)\left(\frac{nm(m+1)}2+1\right)(k+h).
\end{equation}
\end{theorem}

\begin{proof}
Choose a common lattice unit $\ell$.  We first bound the total positive
aggregate difference, $D_+:=d(P)$ for $P:=\{e:d_e>0\}$.  If $D_+=0$,
the equality $d(E)=0$ forces $d=0$.  Otherwise
$\sum_i d_i(P)=D_+$, so some player $i$ satisfies
$\delta_i:=d_i(P)\ge D_+/n$.

By Lemma~\ref{lem:natural-matroid}, lift $x_i^k/\ell$ and $x_i^h/\ell$
to bases $A,B$ with maximum overlap in each fiber.  Contract $A\cap B$
and delete elements outside $A\cup B$.  For each resource $e$, the remaining
disjoint bases $A^\circ,B^\circ$ contain respectively
$(d_{i,e})_+/\ell$ and $(-d_{i,e})_+/\ell$ copies of $e$.  In particular, no
fiber meets both bases.

Apply Corollary~\ref{cor:support-restricted-strong-exchange} to the set of
remaining copies of resources in $E\setminus P$.  Its required imbalance
is positive:
since $d_i(E)=0$,
\[
 |B^\circ\cap C(E\setminus P)|-|A^\circ\cap C(E\setminus P)|
 =-\frac{d_i(E\setminus P)}{\ell}=\frac{\delta_i}{\ell}>0.
\]
The corollary gives $U\subseteq A^\circ\cap C(P)$ and
$V\subseteq B^\circ\cap C(E\setminus P)$, with
$L:=|U|=|V|\ge\delta_i/\ell$.  Reattaching the contracted copies shows
that $A-U+V$ and $B-V+U$ are bases of the original natural matroid.

The construction in Section~\ref{sec:feedback} therefore supplies perfect
matchings in both directions of the packet exchange graph.  The displayed
fiber counts also verify the hypotheses of Lemma~\ref{lem:bad-count}(i),
with $\widehat y_i=x_i^h$ and $\eta=0$, giving
$|Z|\le mt/\ell$.  Lemmas~\ref{lem:bad-cycle}
and~\ref{lem:partial-feedback-path} therefore yield
\[
 \frac{\delta_i}{\ell}\le L\le\frac{m+1}{2}|Z|
 \le\frac{m(m+1)t}{2\ell}.
\]
Multiplying by $\ell$ and using $D_+\le n\delta_i$ gives
$D_+\le nm(m+1)t/2$.  Each positive coordinate is at most $D_+$, and
$d(E)=0$ implies $\sum_{e\notin P}(-d_e)=D_+$, so the same bound holds
for the absolute value of each negative coordinate.  This proves
\eqref{eq:aggregate-proximity}.

For the individual bound, put
$G:=nm(m+1)t/2$, so the aggregate bound gives $|d_r|\le G$ for every $r$.
Fix $i,e$ with $d_{i,e}>0$, repeat the lifting and minor construction for
this player, and apply the exchange corollary to $E\setminus\{e\}$.
Now every copy in $U$ must represent resource $e$, and
$A^\circ$ contains exactly $d_{i,e}/\ell$ such copies.  The corollary gives the
opposite inequality $|U|\ge d_{i,e}/\ell$, so $U$ contains all of them
and $|V|=|U|=d_{i,e}/\ell$.
There are at most $m-1$ resources supporting $V$.  One of them, say $r$,
therefore has at least $|V|/(m-1)$ copies.  Since $B^\circ$ has only
$(-d_{i,r})_+/\ell$ copies of $r$, we obtain
\[
 -d_{i,r}\ge\ell|V_r|\ge\frac{d_{i,e}}{m-1}.
\]
Every copy in $V_r$ has a predecessor in the forward perfect matching
and a successor in the reverse perfect matching.  Both of these matched
copies belong to $U$ and hence represent resource $e$.  Thus the packet moves
$e\to r$ at $x^k$ and $r\to e$ at $x^h$ are feasible.  Because edges
depend only on resources, any chosen $u\in U$ and $v\in V_r$ form a
directed two-cycle.
If $d_{i,e}>(m-1)(G+t)$, then
\[
 s_{i,e}>-G+(m-1)(G+t)\ge t,
 \qquad
 s_{i,r}\le G-\frac{d_{i,e}}{m-1}<-t.
\]
The resulting two-cycle has no exceptional copy, contradicting
Lemma~\ref{lem:bad-cycle}.  This proves
$d_{i,e}\le(m-1)(G+t)$.  Interchanging the two games handles negative
differences and gives \eqref{eq:individual-proximity}.
\end{proof}

\begin{corollary}[Diameter of the packet-equilibrium set]
\label{cor:equilibrium-lattice-diameter}
For two equilibria of the same $k$-integral game,
\begin{align*}
  \max_{e\in E}|x_e-y_e|
  &\le nm(m+1)k,\\
  \max_{\substack{i\in N\\e\in E}}|x_{i,e}-y_{i,e}|
  &\le2(m-1)\left(\frac{nm(m+1)}2+1\right)k.
\end{align*}
\end{corollary}

\begin{proof}
Apply Theorem~\ref{thm:mixed-proximity} to $x$ and $y$ with $h=k$.
The theorem allows any pair of equilibria, including distinct equilibria
at the same packet size.  Substituting $k+h=2k$ gives both estimates.
\end{proof}

\subsection{Proximity to the continuous equilibrium}
\label{sec:continuous-proximity}

Letting the packet size tend to zero turns
Theorem~\ref{thm:mixed-proximity} into a bound on the distance to the
continuous equilibrium.  We first establish lattice approximation and
convergence of packet equilibria.

\begin{lemma}[Rounding to a compatible lattice]
\label{lem:lattice-density}
Let $f/\ell$ be an integral polymatroid rank and let $z\in B(f)$.
There is a point $\widehat z\in B(f)\cap(\ell\Z)^E$ with
$\|\widehat z-z\|_\infty\le\ell$.
\end{lemma}

\begin{proof}
Set $v:=z/\ell\in B(f/\ell)$.  The integral base polytope $B(f/\ell)$
is box-integral; see Fujishige \cite[Sections~3.1--3.2]{Fujishige05}.
This means that intersecting it with any box whose bounds are integers
gives an integral polytope.  In particular,
\[
 K:=B(f/\ell)\cap
 \{q\in\R^E:\lfloor v_e\rfloor\le q_e\le\lceil v_e\rceil
                   \text{ for every }e\}
\]
is nonempty, since it contains $v$, and has an integral vertex $q$.
Set $\widehat z:=\ell q$.  Membership in $B(f/\ell)$ gives
$\widehat z\in B(f)$, while $q\in\Z^E$ gives the required lattice
condition.  Finally, each coordinate of $q$ lies between the floor and
ceiling of $v_e$, so $|\widehat z_e-z_e|=\ell|q_e-v_e|\le\ell$.
\end{proof}

\begin{lemma}[Discrete-to-continuous convergence]
\label{lem:discrete-convergence}
Let $h_j\downarrow0$ be compatible packet sizes and let $x^{h_j}$ be an
equilibrium of the $h_j$-integral game.  Then $x^{h_j}$ converges to the
continuous equilibrium $x^*$.
\end{lemma}

\begin{proof}
The product of the base polytopes is compact, so every subsequence has a
further subsequence converging to some profile $\bar x$.  Relabel this
convergent subsequence as $(x^{h_j})_j$.  Fix a player $i$
and any continuous strategy $z_i\in B(\rho_i)$.  By
Lemma~\ref{lem:lattice-density}, choose
$z_i^{h_j}\in B_{h_j}(\rho_i)$ with $z_i^{h_j}\to z_i$.  The discrete
best-response inequality is
\[
  C_i(x_i^{h_j},x_{-i}^{h_j})
  \le C_i(z_i^{h_j},x_{-i}^{h_j}).
\]
Continuity of $C_i$ lets us pass to the limit and obtain
\[
  C_i(\bar x_i,\bar x_{-i})
  \le C_i(z_i,\bar x_{-i}).
\]
Since $i$ and $z_i$ were arbitrary, $\bar x$ is a continuous equilibrium.
Theorem~\ref{thm:uniqueness} identifies every cluster point with the same
equilibrium $x^*$.  Compactness then implies convergence of the original
sequence: otherwise a subsequence staying a fixed positive distance from
$x^*$ would have a cluster point different from $x^*$.
\end{proof}

We can now pass to the limit in the mixed-packet proximity estimates.

\begin{corollary}[Distance to the continuous equilibrium]
\label{cor:continuous-proximity}
For every packet size $k$ compatible with every player-specific rank
function, and every $k$-integral equilibrium $x^k$,
\begin{align}
  \max_{e\in E}|x_e^k-x_e^*|
  &\le\frac{nm(m+1)}2k,
    \label{eq:continuous-aggregate}\\
  \max_{\substack{i\in N\\e\in E}}|x_{i,e}^k-x_{i,e}^*|
  &\le(m-1)\left(\frac{nm(m+1)}2+1\right)k.
    \label{eq:continuous-individual}
\end{align}
\end{corollary}

\begin{proof}
Fix $k$ and $x^k$ as in the statement and set $h_j:=k/2^j$.
Each $h_j$ is compatible with every rank, since
$\rho_i(S)/h_j=2^j\rho_i(S)/k$ is integral.  Choose an equilibrium
$x^{h_j}$ for every $j$; Lemma~\ref{lem:discrete-convergence} gives
$x^{h_j}\to x^*$.
Theorem~\ref{thm:mixed-proximity} yields, for every resource $e$,
\[
 |x_e^k-x_e^{h_j}|\le\frac{nm(m+1)}2(k+h_j).
\]
Letting $j\to\infty$ proves \eqref{eq:continuous-aggregate}, because
$x_e^{h_j}=\sum_i x_{i,e}^{h_j}\to x_e^*$.  Passing to the limit in
\eqref{eq:individual-proximity} in exactly the same way gives
\eqref{eq:continuous-individual}.
\end{proof}

\subsection{General proximity under changes of the polymatroid rank}
\label{sec:mixed-rank}

We now extend the comparison to games in which both the packet size and the
player's polymatroid rank may change.  This is the situation created
by the scaling construction in Section~\ref{sec:rounding-scaling}.
Let $f_i$ and $g_i$ be two polymatroid rank functions on the
same resource set for every player $i\in N$.  We allow different base values
$f_i(E)$ and $g_i(E)$.  Let $k,h>0$ and assume, for every
$i\in N$, that $f_i$ is compatible with $k$ and $g_i$ is compatible with
$h$.  Choose a common lattice unit $\ell>0$ for $k,h$ and all rank values,
so that
\[
 k/\ell,h/\ell\in\mathbb Z_{>0},
 \qquad
 f_i(S)/\ell,g_i(S)/\ell\in\mathbb Z
 \quad(i\in N, S\subseteq E).
\]
An equilibrium of the $k$-game with strategy spaces $(B(f_i))_{i\in N}$ and
one of the $h$-game with strategy spaces $(B(g_i))_{i\in N}$ both exist by
the integral-polymatroid equilibrium result of Harks, Klimm, and Peis
\cite{HarksKP18}.  Their equilibrium conditions are the corresponding
one-packet move inequalities \eqref{eq:packet-optimality}.

The two bases must now be lifted to the same natural matroid despite the
different ranks.  Lemma~\ref{lem:base-transfer} supplies an auxiliary vector
$\widehat y_i\in B(f_i)\cap(\ell\Z)^E$ with
\[
 \|\widehat y_i-y_i\|_\infty\le2\Delta_i+\ell,
 \qquad \Delta_i:=\max_{S\subseteq E}|f_i(S)-g_i(S)|.
\]
We lift $x_i/\ell$ and $\widehat y_i/\ell$ to bases $A,B$ in the natural
matroid of $f_i/\ell$.  A basis exchange $B-v+u$, with $v\in C_r$ and
$u\in C_e$, makes
$\widehat y_i+\ell(\1_e-\1_r)$ feasible in $B(f_i)$.  The corresponding
reverse edge at the actual equilibrium $y$ requires the full move
$y_i+h(\1_e-\1_r)$ to be feasible in $B(g_i)$.
The forward matching still gives feasible $k$-packet moves at $x_i$.
For the reverse direction, Lemma~\ref{lem:rank-transfer-matching} bounds
how many copies can remain unmatched when edges are required to represent
feasible moves at $y_i$.  The cycle argument continues to use marginal
costs at the actual equilibria $x,y$.  The path-counting bound in
Lemma~\ref{lem:partial-feedback-path} then turns the matching deficiency
and the number of exceptional copies into a bound on the exchanged mass.
This is the link between the rank perturbation and equilibrium proximity.

\begin{lemma}[Transfer between nearby base polytopes]
\label{lem:base-transfer}
Let $f/\ell$ and $g/\ell$ be integral polymatroid rank functions and set
\[
 \Delta(f,g):=\max_{S\subseteq E}|f(S)-g(S)|.
\]
For every $y\in B(g)\cap(\ell\mathbb Z)^E$ there is a vector
$\widehat y\in B(f)\cap(\ell\mathbb Z)^E$ satisfying
\begin{equation}
\label{eq:base-transfer-bound}
 \|\widehat y-y\|_\infty\le 2\Delta(f,g)+\ell.
\end{equation}
\end{lemma}

\begin{proof}
For a permutation $\pi=(e_1,\ldots,e_m)$, put
$S_j^\pi:=\{e_1,\ldots,e_j\}$ and $S_0^\pi:=\varnothing$.
By the greedy construction in \eqref{eq:greedy-base}, the corresponding vertices
$v^f(\pi)$ and $v^g(\pi)$ have coordinates
$f(S_j^\pi)-f(S_{j-1}^\pi)$ and $g(S_j^\pi)-g(S_{j-1}^\pi)$.
The two rank differences in each coordinate give
\[
 |v^f(\pi)_{e_j}-v^g(\pi)_{e_j}|
 \le |f(S_j^\pi)-g(S_j^\pi)|
     +|f(S_{j-1}^\pi)-g(S_{j-1}^\pi)|
 \le2\Delta(f,g).
\]
Every point of $B(g)$ is a convex combination of its greedy vertices, so write
\[
 y=\sum_\pi\lambda_\pi v^g(\pi),\qquad
 \lambda_\pi\ge0,\quad \sum_\pi\lambda_\pi=1.
\]
Using the same coefficients, set $z:=\sum_\pi\lambda_\pi v^f(\pi)$.
Convexity gives $z\in B(f)$, and
\[
 \|z-y\|_\infty
 \le\sum_\pi\lambda_\pi\|v^f(\pi)-v^g(\pi)\|_\infty
 \le2\Delta(f,g).
\]
To obtain a point on the $\ell$-lattice, apply
Lemma~\ref{lem:lattice-density} and choose
$\widehat y\in B(f)\cap(\ell\Z)^E$ with
$\|\widehat y-z\|_\infty\le\ell$.  The triangle inequality now gives
\eqref{eq:base-transfer-bound}.
\end{proof}

The auxiliary vector $\widehat y$ allows us to construct basis exchanges
in the natural matroid of $f/\ell$.  A corresponding reverse move from
resource $r$ to resource $s$ can fail at $y$ for two reasons: either
$y_r<h$, or a constraint containing $s$ and excluding $r$ has slack less
than $h$.  The first case accounts for a bounded number of source copies.
For the second, we combine the obstructing constraints into one set with
bounded slack and use Hall's theorem to control the unmatched copies.
The next lemma carries out these steps.

\begin{lemma}[Reverse matching after a rank transfer]
\label{lem:rank-transfer-matching}
Let $y\in B(g)\cap(\ell\mathbb Z)^E$ and
$\widehat y\in B(f)\cap(\ell\mathbb Z)^E$ satisfy
$\|y-\widehat y\|_\infty\le\eta$.  In the natural matroid of $f/\ell$,
choose a basis $B$ satisfying $\pi(B)=\widehat y/\ell$.  Suppose that disjoint sets of copies
$V\subseteq B$ and $U\cap B=\varnothing$ have equal size $L$, no resource
fiber contains a member of both $U$ and $V$, and
$B-V+U\in\mathcal B(\mathcal M_{f/\ell})$.
For a copy $v\in V$ of resource $r$ and a copy $u\in U$ of resource $s$,
include the directed edge $v\to u$ when
\[
 y+h(\mathbf 1_s-\mathbf 1_r)\in B(g).
\]
If $\Delta=\Delta(f,g)$ and $H_m=m+\lfloor m^2/4\rfloor$, this graph has a matching of size
at least
\begin{equation}
\label{eq:rank-transfer-deficiency}
 L-\frac{H_mh+\Delta+2m\eta}{\ell}.
\end{equation}
\end{lemma}

\begin{proof}
\emph{Source fibers with less than one packet.}
Discard all copies in a source fiber $C_r$ with $y_r<h$.  Since
$\widehat y_r\le y_r+\eta<h+\eta$, fewer than
$m(h+\eta)/\ell$ copies are discarded.  Let $V'$ be the remaining source
set.  We bound the Hall deficiency from $V'$ into $U$: it is enough to bound
$|X|-|N(X)|$ for every $X\subseteq V'$, where $N(X)$ denotes its neighbors
in $U$.

\emph{Constraints obstructing reverse moves.}
Fix $X\subseteq V'$, let $Q$ be the set of resources whose fibers meet $X$,
and put $W=U\setminus N(X)$.  Let $F$ be the resources whose fibers meet
$W$.  If $Q$ is empty, then $X=\varnothing$; if $F$ is empty, then $N(X)=U$
and $|X|\le|U|$.  In both cases the deficiency is nonpositive.  Otherwise,
for every $r\in Q$ and $s\in F$ the full $h$-exchange from $r$ to $s$ is
infeasible at $y$, while $y_r\ge h$.  The exchange-capacity formula
\eqref{eq:capacity-formula}, applied to $g$, gives a set $S_{r,s}$ with
\[
 s\in S_{r,s},\qquad r\notin S_{r,s},\qquad
 g(S_{r,s})-y(S_{r,s})<h.
\]
\emph{Combining the obstructing constraints.}
For every $s\in F$, put
\[
 T_s:=\bigcap_{r\in Q}S_{r,s}.
\]
It contains $s$ and avoids all of $Q$.  To control its slack, write
$a(S):=g(S)-y(S)$.  Since $y\in B(g)$, this is a nonnegative submodular
function.  For any two sets, submodularity and nonnegativity imply
\[
 a(S\cap S')\le a(S)+a(S'),\qquad
 a(S\cup S')\le a(S)+a(S').
\]
Repeatedly applying the intersection inequality gives
\[
 g(T_s)-y(T_s)
 \le\sum_{r\in Q}\bigl(g(S_{r,s})-y(S_{r,s})\bigr)
 <|Q|h.
\]
Now set $T:=\bigcup_{s\in F}T_s$.  A second application of the same
submodular union inequality shows that $T$ contains $F$, avoids $Q$, and
satisfies
\begin{equation}
\label{eq:rank-transfer-near-tight}
 g(T)-y(T)<|F||Q|h\le \lfloor m^2/4\rfloor h.
\end{equation}
For the last bound, the disjoint fiber supports give $F\cap Q=\varnothing$,
so $|F|+|Q|\le m$ and $|F||Q|\le\lfloor m^2/4\rfloor$.

Thus one set contains every resource represented among the nonneighbors
and avoids every source resource in $X$, with slack less than $\lfloor m^2/4\rfloor h$.

\emph{Bounding the number of unmatched copies.}
Let $u,v\in(\ell\mathbb Z)^E$ be $\ell$ times the count projections of
$U,V$, respectively.
Feasibility of $\widehat y-v+u$ in $B(f)$ gives
\[
 u(T)-v(T)\le f(T)-\widehat y(T).
\]
The right-hand side is at most
\begin{align*}
 f(T)-\widehat y(T)
 &\le g(T)-y(T)+|f(T)-g(T)|+|y(T)-\widehat y(T)|\\
 &<\lfloor m^2/4\rfloor h+\Delta+m\eta.
\end{align*}
The three terms account for the slack in $B(g)$, the rank perturbation,
and the displacement of the auxiliary vector, respectively.
Every copy in $W$ contributes to $u(T)$, and $T$ avoids every fiber
represented in $X$; hence $v(T)\le\ell|V\setminus X|$.  Therefore
\[
 \ell\bigl(|W|-|V\setminus X|\bigr)
 <\lfloor m^2/4\rfloor h+\Delta+m\eta.
\]
Since $|U|=|V|=L$ and $W=U\setminus N(X)$, the left side equals
\[
 \ell\bigl((L-|N(X)|)-(L-|X|)\bigr)
 =\ell(|X|-|N(X)|).
\]
This holds for every $X\subseteq V'$.  Hall's deficiency theorem now gives a matching from
$V'$ into $U$ that misses at most
$(\lfloor m^2/4\rfloor h+\Delta+m\eta)/\ell$ vertices of $V'$.  Adding the discarded source
copies yields a total deficiency of at most
\[
 \frac{m(h+\eta)+\lfloor m^2/4\rfloor h+\Delta+m\eta}{\ell}
 =\frac{H_mh+\Delta+2m\eta}{\ell},
\]
which is \eqref{eq:rank-transfer-deficiency}.
\end{proof}

The resulting proximity bounds are linear in $k+h+\Delta$, with
polynomial dependence on $n,m$ and no dependence on cost coefficients.
Write $G_{n,m}(k,h,\Delta;\ell)$ and $J_{n,m}(k,h,\Delta;\ell)$ for the
aggregate and individual bounds.  With $t=k+h$ as above, define them
using the abbreviations $G_{n,m}$ and $J_{n,m}$ by
\begin{align}
 k_{\max}&:=\max\{k,h\},\qquad
 \Delta:=\max_{\substack{i\in N\\S\subseteq E}}|f_i(S)-g_i(S)|,
 \label{eq:mixed-rank-basic-constants}\\
 \eta&:=2\Delta+\ell,\notag\\
 G_{n,m}
 &:=n\left[
 m\eta+\frac m2\bigl(H_mk_{\max}+\Delta+2m\eta\bigr)
 +\frac{m(m+1)}2(t+\eta)\right],
 \label{eq:mixed-rank-G}\\
 J_{n,m}
 &:=\eta+\frac m2\bigl(H_mk_{\max}+\Delta+2m\eta\bigr)\notag\\
 &\quad+\frac{m+1}{2}
   \bigl((m-1)(G_{n,m}+t+\eta)\bigr).
 \label{eq:mixed-rank-J}
\end{align}
Separating the packet-size and rank-perturbation terms gives, with absolute
implied constants,
\[
 G_{n,m}=O\bigl(nm^3(k+h)+nm^2\Delta\bigr),\qquad
 J_{n,m}=O\bigl(nm^5(k+h)+nm^4\Delta\bigr).
\]
In particular, at a halving step with $h=k/2$ and
$\Delta\le(m^2+1)k$, the individual bound is $O(nm^6k)$.
These are functions of both packet sizes, the rank distance, and the
common lattice unit.  Since $\ell\le\min\{k,h\}$, its contribution is
absorbed by $k+h$ in the preceding order bounds.

\begin{theorem}[Mixed-rank equilibrium proximity]
\label{thm:mixed-rank-proximity}
Let $x$ be an equilibrium of the $k$-integral game with
strategy spaces $(B(f_i))_{i\in N}$ and let $y$ be an equilibrium of the
$h$-integral game with strategy spaces $(B(g_i))_{i\in N}$.  Assume the
affine costs \eqref{eq:affine-specialization}; for every $i\in N$, assume
that $f_i$ is compatible with $k$ and $g_i$ is compatible with $h$.  Let
$\ell$ and the bound functions be as defined above.  Then
\begin{align}
 \max_{e\in E}|x_e-y_e|&\le G_{n,m}(k,h,\Delta;\ell),
   \label{eq:mixed-rank-aggregate}\\
 \max_{\substack{i\in N\\e\in E}}|x_{i,e}-y_{i,e}|
 &\le J_{n,m}(k,h,\Delta;\ell).
   \label{eq:mixed-rank-individual}
\end{align}
The bounds are independent of all cost coefficients and of the absolute rank
magnitudes; the rank data enter only through the displayed distance $\Delta$
and the common lattice unit $\ell$.
\end{theorem}

\begin{proof}
Use the differences from Section~\ref{sec:feedback}.
Set
\[
 \beta_*:=\frac{H_mk_{\max}+\Delta+2m\eta}{\ell}.
\]
For any player $i$, Lemma~\ref{lem:base-transfer} supplies an auxiliary vector
$\widehat y_i\in B(f_i)\cap(\ell\Z)^E$ with
$\|\widehat y_i-y_i\|_\infty\le\eta$.  Choose maximum-overlap lifts $A,B$
of $x_i/\ell,\widehat y_i/\ell$ in the natural matroid of $f_i/\ell$.
After contracting their common copies, the remaining fiber counts are
$(x_{i,e}-\widehat y_{i,e})_+/\ell$ on the $A$-side and
$(\widehat y_{i,e}-x_{i,e})_+/\ell$ on the $B$-side.  Any exchange
between these remaining copies satisfies the multiplicity assumptions of
Lemma~\ref{lem:bad-count}.

For every simultaneous exchange $U,V$, the forward matching is perfect,
as in Section~\ref{sec:feedback}.  For the reverse direction,
Lemma~\ref{lem:rank-transfer-matching} supplies a matching of feasible
packet moves at $y_i$ whose deficiency is at most
\[
 \frac{H_mh+\Delta+2m\eta}{\ell}\le\beta_*.
\]
We apply Lemma~\ref{lem:partial-feedback-path} with
$\beta^+=0$ and $\beta^-=\beta_*$.  The use of $k_{\max}$ makes this
bound valid for the interchanged games used below.

For the aggregate bound, put $P:=\{e:d_e>0\}$ and $D_+:=d(P)$.
If $D_+>0$, choose $i$ with $\delta_i:=d_i(P)\ge D_+/n$.
When $\delta_i\le m\eta$, we already have $D_+\le nm\eta\le G_{n,m}$.
Otherwise, replace $y_i\in B(g_i)$ by the auxiliary vector
$\widehat y_i\in B(f_i)$ for the exchange construction.  The imbalance
$x_i(P)-y_i(P)$ changes by at most $|P|\eta$, since
$\|\widehat y_i-y_i\|_\infty\le\eta$.  Thus
\[
 \widetilde\delta_i:=x_i(P)-\widehat y_i(P)
 =\delta_i+y_i(P)-\widehat y_i(P)
 \ge\delta_i-|P|\eta\ge\delta_i-m\eta>0.
\]
The vectors $x_i$ and $\widehat y_i$ have the same total $f_i(E)$.
Among the copies of resources in $E\setminus P$, the $B$-side has
$\widetilde\delta_i/\ell$ more copies than the $A$-side.  Contract common copies and delete elements
outside $A\cup B$, as in the proof of
Theorem~\ref{thm:mixed-proximity}, and apply
Corollary~\ref{cor:support-restricted-strong-exchange} to the copies of
resources in $E\setminus P$.  It gives $U\subseteq C(P)$ and
$V\subseteq C(E\setminus P)$, with
$L:=|U|=|V|\ge\widetilde\delta_i/\ell\ge(\delta_i-m\eta)/\ell$.
Lemmas~\ref{lem:bad-count}(i), \ref{lem:bad-cycle},
and~\ref{lem:partial-feedback-path} give
\[
 \frac{\delta_i-m\eta}{\ell}\le L
 \le\frac m2\beta_*+\frac{m(m+1)}{2\ell}(t+\eta).
\]
Multiplying by $\ell$ and rearranging gives
\[
 \delta_i\le m\eta+
 \frac m2\bigl(H_mk_{\max}+\Delta+2m\eta\bigr)
 +\frac{m(m+1)}2(t+\eta)=G_{n,m}/n.
\]
Thus $D_+\le n\delta_i\le G_{n,m}$; the case $D_+=0$ is immediate.
Since $x,y$ may have different total allocations, we bound the negative
aggregate differences by interchanging the two games and repeating the
argument.  The constants are unchanged, and this bounds $\sum_{e:d_e<0}(-d_e)$ by $G_{n,m}$ as well.
Each coordinate difference is bounded by the corresponding sum, proving
\eqref{eq:mixed-rank-aggregate}.

For the individual bound, fix $i,e$ with $d_{i,e}>0$.  If
$d_{i,e}\le G_{n,m}+t$, the claimed bound follows directly from
\eqref{eq:mixed-rank-J}.  Otherwise $G_{n,m}\ge nm\eta$ gives
$d_{i,e}>\eta$, so $x_{i,e}>\widehat y_{i,e}$.
Apply Corollary~\ref{cor:support-restricted-strong-exchange} in the same
minor to the copies of resources in $E\setminus\{e\}$.  It requires
$|U|\ge(x_{i,e}-\widehat y_{i,e})/\ell$, while all copies available to
$U$ represent resource $e$ and number exactly this quantity.  Equality holds,
and
\[
 \ell L=x_{i,e}-\widehat y_{i,e}\ge d_{i,e}-\eta.
\]
Lemma~\ref{lem:bad-count}(ii) gives
$|Z|\le(m-1)(G_{n,m}+t+\eta)/\ell$.
Since every directed cycle meets $Z$, the matching deficiencies and
Lemma~\ref{lem:partial-feedback-path} give
$L\le m\beta_*/2+(m+1)|Z|/2$.  Combining this with
$d_{i,e}\le\eta+\ell L$ yields
\[
 d_{i,e}\le\eta+\frac{m\ell}{2}\beta_*
 +\frac{m+1}{2}(m-1)(G_{n,m}+t+\eta)=J_{n,m}.
\]
Interchanging the games handles $d_{i,e}<0$.
\end{proof}

\begin{corollary}[Continuous rank sensitivity]
\label{cor:continuous-rank-sensitivity}
Consider two games on the same player and resource sets, with the same
affine costs \eqref{eq:affine-specialization} and arbitrary real polymatroid
ranks $(f_i)_{i\in N}$ and $(g_i)_{i\in N}$.  Put
$\Delta=\max_{i,S}|f_i(S)-g_i(S)|$.  Their unique continuous equilibria
$x^*$ and $y^*$ satisfy
\[
 \max_{e\in E}|x_e^*-y_e^*|=O(nm^2\Delta),\qquad
 \max_{\substack{i\in N\\e\in E}}|x_{i,e}^*-y_{i,e}^*|
 =O(nm^4\Delta).
\]
The implied constants are universal and independent of the cost
coefficients and the absolute rank magnitudes.
\end{corollary}
\begin{proof}
First suppose that all ranks are rational.  Choose a positive integer
$D$ such that every rank value of every $f_i$ and $g_i$ is an integer
multiple of $1/D$.  Such a common denominator
exists because there are finitely many rational rank values.
For $j\ge0$, put $q_j=1/(D2^j)$ and choose equilibria $x^{q_j}$ and
$y^{q_j}$ of the respective $q_j$-integral games.  These exist by the
integral-polymatroid equilibrium theorem of Harks, Klimm, and
Peis~\cite{HarksKP18} after rescaling the packet size.
Lemma~\ref{lem:discrete-convergence} gives $x^{q_j}\to x^*$ and
$y^{q_j}\to y^*$.

Apply Theorem~\ref{thm:mixed-rank-proximity} with
$k=h=\ell=q_j$.  The explicit bounds
\eqref{eq:mixed-rank-G}--\eqref{eq:mixed-rank-J} give universal constants
$C_G,C_J$ such that, for every $j$,
\begin{align*}
 \max_e|x_e^{q_j}-y_e^{q_j}|
 &\le C_G\bigl(2nm^3q_j+nm^2\Delta\bigr),\\
 \max_{i,e}|x_{i,e}^{q_j}-y_{i,e}^{q_j}|
 &\le C_J\bigl(2nm^5q_j+nm^4\Delta\bigr).
\end{align*}
Letting $j\to\infty$ eliminates the packet-size terms and proves the
claims for rational ranks, with constants independent of costs and rank
magnitudes.

Now let the ranks be real.  The normalized, nondecreasing, submodular
functions on the finite set $E$ form a rational polyhedral cone.  Rational
points are dense in this cone, so choose rational polymatroid ranks
$f_i^{(r)}\to f_i$ and $g_i^{(r)}\to g_i$ for every player.  Their maximum
rank distance $\Delta_r$ converges to $\Delta$.  We show that their
continuous equilibria converge to those of the limiting games.

Let $x^{(r)}$ be the equilibrium for $(f_i^{(r)})_i$.  These profiles are
uniformly bounded by the convergent total ranks.  Any convergent
subsequence has a feasible limit $\bar x$, by passing to the limit in
all base constraints.  Fix a player $i$ and a deviation $z_i\in B(f_i)$.
Express $z_i$ as a convex combination of the greedy vertices of $B(f_i)$.
For each ordering, the greedy vertex for $f_i^{(r)}$ converges to the one
for $f_i$, since its coordinates are rank increments.  Using the same
convex coefficients gives $z_i^{(r)}\in B(f_i^{(r)})$ with
$z_i^{(r)}\to z_i$.  Passing to the limit in
\[
 C_i(x_i^{(r)},x_{-i}^{(r)})
 \le C_i(z_i^{(r)},x_{-i}^{(r)})
\]
shows that $\bar x$ is an equilibrium for $(f_i)_i$.
Uniqueness (Theorem~\ref{thm:uniqueness}) implies $x^{(r)}\to x^*$.
The same argument gives $y^{(r)}\to y^*$ for the ranks $(g_i^{(r)})_i$.
Apply the rational-rank bounds to $x^{(r)},y^{(r)}$ and let $r\to\infty$.
Since $\Delta_r\to\Delta$, both bounds follow with the same constants.
\end{proof}

\section{Compatible Submodular Rounding and Packet Scaling}
\label{sec:rounding-scaling}

Our outer scaling architecture follows the packet-halving method of Harks and
Timmermans
\cite{HarksT22,HarksT22Correction}.  In their
singleton setting, after
halving the packet size a \hyperlink{cite.HarksT22}{\textsc{Restore}} routine re-establishes equilibrium
and an \hyperlink{cite.HarksT22}{\textsc{Add}} routine schedules the remaining demand; proximity bounds
control both operations.  Here compatible submodular rounding replaces
singleton demand rounding, and mixed-rank proximity permits a single residual
$\RankRestore$ call to absorb both the scale change and the change in
rounded base value.

We first construct a compatible rank at every dyadic scale.  Each rounded-rank value uses one query to the original rank oracle after
the base value has been queried once.

\paragraph{Why direct rounding fails.}
Take $E=\{1,2,3,4\}$ and the modular rank $\rho(S)=|S|$.
For packet size $k=2$, direct upward rounding gives
$R_2(S)=2\lceil|S|/2\rceil$.
With $A=\{1,2\}$ and $B=\{2,3\}$, we obtain
\[
 R_2(A)+R_2(B)=2+2=4
 <4+2=R_2(A\cup B)+R_2(A\cap B).
\]
Thus submodularity fails even for a modular rank.  Truncating these values
at the rounded total demand $d_2=4$ leaves the violation unchanged.
Direct downward rounding also fails: the two singleton values for
$\{1\}$ and $\{2\}$ are zero, whereas their union has rounded value two.

\paragraph{A correction that preserves submodularity.}
For an integer $m=|E|$, define the cardinality correction
\begin{equation}
\label{eq:psi-definition}
 \Psi_m(S):=|S|(2m-|S|).
\end{equation}
For a coarse packet size $k>1$, put
\begin{equation}
\label{eq:coarse-rounding}
 R_k(S):=k\left\lceil\frac{\rho(S)}k\right\rceil,
 \qquad
 \overline\rho_k(S):=R_k(S)+k\Psi_m(S).
\end{equation}
For the sets $A=\{1,2\}$ and $B=\{2,3\}$ in the example, the correction
$2\Psi_4$ contributes submodular slack four, repairing the rounding deficit
of two.  In general, its slack is $2k|A\setminus B||B\setminus A|$,
as shown in \eqref{eq:psi-slack}.  This is at least $2k$ for every
nonnested pair and absorbs the rounding error; nested pairs satisfy the
submodular inequality with equality automatically.

For a fine dyadic packet size $0<k\le1$, put
$\overline\rho_k:=\rho$.  Finally define the rounded base value and the
effective rank
\begin{equation}
\label{eq:effective-rounding}
 d_k:=k\left\lfloor\frac{\rho(E)}k\right\rfloor,
 \qquad
 \sigma_k(S):=\min\{\overline\rho_k(S),d_k\}.
\end{equation}
At fine scales we only use $k=2^{-q}$, so integrality of $\rho$ implies
$\rho(S)/k=2^q\rho(S)\in\mathbb Z$ for every $S\subseteq E$;
hence compatibility is automatic throughout the fine dyadic chain, while
$d_k=\rho(E)$ and $\sigma_k=\rho$.

\begin{theorem}[Compatible submodular rounding]
\label{thm:compatible-rounding}
Let $\rho$ be an integral polymatroid rank on an $m$-element ground set.
For every dyadic packet size $k=2^j$, $j\in\Z$, the function $\sigma_k$ in
\eqref{eq:effective-rounding} is a polymatroid rank, all its values are
multiples of $k$, and $\sigma_k(E)=d_k$.  Moreover:
\begin{enumerate}[label=\textnormal{(\roman*)},leftmargin=2.6em]
 \item if $k\le1$, then $\sigma_k=\rho$;
 \item if $k>1$, then
 \begin{equation}
 \label{eq:rounding-approximation}
  |\sigma_k(S)-\rho(S)|<(m^2+1)k
  \qquad(S\subseteq E);
 \end{equation}
 \item for $h=k/2$,
 \begin{equation}
 \label{eq:successive-rank-distance}
  \max_{S\subseteq E}|\sigma_k(S)-\sigma_h(S)|
  \le(m^2+1)k.
 \end{equation}
\end{enumerate}
After one initial query for $\rho(E)$, every value of $\sigma_k$ is obtained
from one further value-oracle query for $\rho$.
\end{theorem}

\begin{proof}
The function $R_k$ is normalized and nondecreasing but need not be
submodular.  The correction $\Psi_m$ is normalized and nondecreasing, since
its increment at cardinality $s<m$ is $2m-2s-1>0$.  With
$p_A:=|A\setminus B|$ and $p_B:=|B\setminus A|$, direct expansion gives
\begin{equation}
\label{eq:psi-slack}
 \Psi_m(A)+\Psi_m(B)-\Psi_m(A\cup B)-\Psi_m(A\cap B)=2p_Ap_B.
\end{equation}
If $A,B$ are nested, the submodularity inequality for $R_k$ is an identity.
If they cross, $p_A,p_B\ge1$.  Each rounding error
$R_k(S)-\rho(S)$ lies in $[0,k)$, so submodularity of $\rho$ gives
\[
 R_k(A)+R_k(B)-R_k(A\cup B)-R_k(A\cap B)>-2k.
\]
The correction in \eqref{eq:psi-slack} is at least $2k$, proving
submodularity of $\overline\rho_k$.

Truncation preserves polymatroid ranks; see Fujishige
\cite[Section~3.1]{Fujishige05}.  The function $\sigma_k$ is a polymatroid
rank.  Both arguments of the minimum in
\eqref{eq:effective-rounding} are multiples of $k$, and
$\overline\rho_k(E)\ge d_k$, proving compatibility and
$\sigma_k(E)=d_k$.  For $k\le1$, the definition and integrality of $\rho$
give $d_k=\rho(E)$, hence $\sigma_k=\rho$.

For $k>1$, $0\le R_k(S)-\rho(S)<k$ and
$0\le\Psi_m(S)\le m^2$.  It follows that
$\overline\rho_k(S)-\rho(S)<(m^2+1)k$.  If truncation makes
$\sigma_k(S)<\rho(S)$, then $\sigma_k(S)=d_k$ and
\[
 0<\rho(S)-\sigma_k(S)
 \le\rho(E)-d_k<k.
\]
This proves \eqref{eq:rounding-approximation}.

It remains to compare consecutive scales.  The elementary inequality
\[
 |\min\{a,c\}-\min\{b,d\}|\le
 \max\{|a-b|,|c-d|\}
\]
reduces the claim to the untruncated rounded ranks and base values.  If $h>1$,
then $k=2h$, $|d_k-d_h|\le h$, and
\begin{align*}
 |\overline\rho_k(S)-\overline\rho_h(S)|
 &\le|R_k(S)-R_h(S)|+(k-h)\Psi_m(S)\\
 &<k+\frac{k}{2}m^2
 \le(m^2+1)k.
\end{align*}
If $k=2$ and $h=1$, then $\sigma_h=\rho$ and
\eqref{eq:successive-rank-distance} follows from
\eqref{eq:rounding-approximation}.  If $k\le1$, both ranks equal $\rho$.
The oracle assertion is immediate from the formulas.
\end{proof}

\subsection{Residual bases and shifted games}

The restore procedure freezes a common lower vector and solves only for the
remaining mass.  We state the required residual operation in oracle form.

\begin{lemma}[Residual polymatroid]
\label{lem:residual-polymatroid-new}
Let $f$ be a polymatroid rank and let $w\in P(f)$.  Define
\begin{equation}
\label{eq:residual-rank-new}
 f^w(S):=\min_{T:\,S\subseteq T\subseteq E}\{f(T)-w(T)\}.
\end{equation}
Then $f^w$ is a polymatroid rank and
\begin{equation}
\label{eq:residual-base-translation-new}
 B(f^w)=\{z\in\mathbb R_{\ge0}^E:w+z\in B(f)\}.
\end{equation}
If $f$ and $w$ are $h$-integral, then $f^w$ is $h$-integral.  A value of
$f^w(S)$ can be found by one submodular-function minimization with all
elements of $S$ forced into the minimizing set.
\end{lemma}

\begin{proof}
The function $a(T):=f(T)-w(T)$ is normalized, nonnegative, and submodular.
Its monotone closure is $a^\uparrow(S)=\min_{T\supseteq S}a(T)=f^w(S)$.
It is nonnegative and equals zero at $\varnothing$, because
$a(\varnothing)=0$ is one of the candidates there.  If $S\subseteq S'$, the
minimum for $S'$ ranges over fewer sets than the minimum for $S$; hence
$f^w(S)\le f^w(S')$.  If $T_A,T_B$ minimize the expressions for $A,B$, then
\begin{align*}
 f^w(A)+f^w(B)
 &=a(T_A)+a(T_B)\\
 &\ge a(T_A\cap T_B)+a(T_A\cup T_B)\\
 &\ge f^w(A\cap B)+f^w(A\cup B),
\end{align*}
so $f^w$ is submodular.  For $z\ge0$, the inequalities
$z(S)\le a(S)$ for all $S$ are equivalent to $z(S)\le a^\uparrow(S)$ for
all $S$: one implication uses $a^\uparrow(S)\le a(S)$, while the other uses
$z(S)\le z(T)\le a(T)$ for every $T\supseteq S$.  Since
$f^w(E)=f(E)-w(E)$, adding the base equality proves
\eqref{eq:residual-base-translation-new}.  Integrality and the oracle claim
follow directly from \eqref{eq:residual-rank-new}.
\end{proof}

Let $g_i$ be polymatroid ranks and let $w=(w_i)_i$ satisfy
$w_i\in P(g_i)$.  Put $w_e=\sum_jw_{j,e}$ and define the shifted costs
on the residual bases $B(g_i^{w_i})$ by
\begin{equation}
\label{eq:shifted-cost-new}
 \widetilde c_{i,e}(z_e)
 :=a_{i,e}z_e+\widetilde b_{i,e},
 \qquad
 \widetilde b_{i,e}:=b_{i,e}+a_{i,e}(w_e+w_{i,e}).
\end{equation}
Expansion gives
\begin{equation}
\label{eq:shifted-equivalence-new}
 C_i(w+z)
 =\sum_ez_{i,e}\widetilde c_{i,e}(z_e)
 +\sum_ew_{i,e}
   \bigl(a_{i,e}(w_e+z_{-i,e})+b_{i,e}\bigr).
\end{equation}
The last sum is independent of $z_i$.  The shifted residual game is
strategically equivalent to the original game restricted to profiles that
coordinatewise dominate $w$.

For the algorithmic statements, a rank-oracle call returns its integer value
in binary, and all rational cost coefficients are given by binary numerators
and denominators.  We use exact rational arithmetic.  Write
$\mathsf{SFM}(m,L)$ for an upper bound on the time needed to minimize an
$m$-element integer-valued submodular function whose oracle values have at
most $L$ bits, including its oracle-call cost.

Let $R_{\max}:=\max_{i\in N}\rho_i(E)$ be the largest original base value.
We use Algorithm~1 of Harks, Klimm, and Peis~\cite{HarksKP18}.
In the published journal version, Theorem~5.1 proves correctness, and
Theorem~5.2 bounds the number
of restricted best-response updates by $n^2mQ^3$.  Their
Theorems~3.2 and~3.4 supply the unit moves and additions.  The cost model in their Section~4 allows
player-specific functions $F_{i,e}(\alpha;\gamma)$ of own usage $\alpha$
and aggregate opponents' usage $\gamma$, subject to the regularity
conditions below.  Our affine congestion costs are a special case.

\begin{proposition}[Integral polymatroid equilibrium algorithm]
\label{prop:hkp-algorithm}
Let $\varphi_i$ be integral polymatroid ranks, and let $\mathrm{dem}_i$ be
nonnegative integers with $\mathrm{dem}_i\le\varphi_i(E)$.  Give player $i$
the strategies
\[
 \{z_i\in P(\varphi_i)\cap\Z^E:z_i(E)=\mathrm{dem}_i\}
\]
and cost $\sum_e F_{i,e}(z_{i,e};z_{-i,e})$.
Assume $F_{i,e}:\Z_{\ge0}^2\to\R_{\ge0}$ is nondecreasing in both
arguments, and for integer counts $\alpha\ge1$, $\gamma\ge0$ define
\[
 F^-_{i,e}(\alpha;\gamma)
 :=F_{i,e}(\alpha;\gamma)-F_{i,e}(\alpha-1;\gamma).
\]
Suppose
\begin{equation}
\label{eq:hkp-regularity}
 F^-_{i,e}(\alpha;\gamma)\le F^-_{i,e}(\alpha;\gamma+1)
 \le F^-_{i,e}(\alpha+1;\gamma)
 \qquad(\alpha\ge1,\ \gamma\ge0).
\end{equation}
Then a pure equilibrium can be found in $O(n^2mQ^3)$ restricted best-response updates,
where $Q:=\max\{1,\max_i\mathrm{dem}_i\}$.  The algorithm, denoted by
$\mathsf{PAlg}$, starts at zero demand, increases demands one unit at a time,
and chooses a best response adding one unit after each demand increase.
Equilibrium is then restored by restricted best responses that move one
unit from the resource whose load most recently increased.  The sensitivity
results ensure that a best response of this form exists whenever an
improving player is selected.  Intermediate strategy spaces are the
corresponding demand truncations of the fixed ranks $\varphi_i$.
\end{proposition}

For our residual game, set $\varphi_i=g_i^{w_i}/h$ and
$\mathrm{dem}_i=g_i^{w_i}(E)/h$.  For one resource, write $a$ for its slope
and $\widetilde b$ for its shifted intercept.  In own and other-player
packet counts $\alpha,\gamma$, its cost and left difference are
\[
 F(\alpha;\gamma)=ah^2\alpha(\alpha+\gamma)+\widetilde b h\alpha,
 \qquad
 F^-(\alpha;\gamma)=ah^2(2\alpha+\gamma-1)+\widetilde b h.
\]
These satisfy \eqref{eq:hkp-regularity}, which also implies discrete
convexity in $\alpha$.  The playerwise latency shift from
Section~\ref{sec:mixed-arcs} is chosen anew in each restore phase,
after the frozen vector determines the shifted intercepts.  It makes
$\widetilde b\ge0$, so $F$ is nonnegative and nondecreasing in both arguments.  The shift preserves best responses at
each intermediate demand, since all strategies then have the same total
allocation.  Proposition~\ref{prop:hkp-algorithm} applies.

For the oracle implementation, enumerate feasible additions for the player
whose demand increases.  During restoration, consider only moves from the
resource that received a packet in the preceding update.  For a selected
improving player, choose a minimum-cost feasible move from that source.
The sensitivity results guarantee that this restricted choice is a best
response at the profiles reached by $\PAlg$.  An addition has at most
$m$ destinations.  During restoration, the source is fixed, so at most
$m-1$ destinations need to be tested for each of the $n$ players.  Including
the scan that detects termination, each update therefore requires
$O(nm)$ candidate feasibility tests and exact cost comparisons.  For any nonnegative candidate $z_i'$, the equivalence used
in the proof of Lemma~\ref{lem:residual-polymatroid-new} gives
\[
 z_i'\in P(g_i^{w_i})
 \quad\Longleftrightarrow\quad
 w_i+z_i'\in P(g_i).
\]
Feasibility is decided by the single submodular minimization
\[
 \min_{S\subseteq E}
   \bigl\{g_i(S)-w_i(S)-z_i'(S)\bigr\}\ge0,
\]
together with nonnegativity and the required total value.  A rounded-rank value
$g_i(S)$ uses one query to the original rank oracle.  Division by the current
packet size makes the displayed submodular function integer-valued.  Each restricted best-response update uses
$O(nm)$ calls to submodular-function minimization and a polynomial number of
exact arithmetic operations.

\subsection{A rank-changing restore lemma}

At a halving step $h=k/2$,
$\RankRestore$ freezes the allocation
$w_{i,e}=\max\{x^k_{i,e}-\tau,0\}$ for a buffer $\tau=\mathfrak B_{n,m}k$.
The residual solver tests only deviations that keep the full allocation
above $w$.  To transfer its equilibrium inequalities to the target game,
we must retain every feasible target packet move.  At a coordinate with
$w_{i,e}=0$, the residual and target allocations agree; at a coordinate
with $w_{i,e}>0$, one full residual packet suffices to allow any target
packet removal.  We therefore choose the buffer to cover the change in
rank and leave one removable packet after the possible equilibrium
displacements.  It suffices to impose
\[
 \tau\ge(m^2+1)k,
 \qquad
 \tau\ge\widehat J_{n,m}k+\Lambda_{n,m}k+h.
\]
Here $\widehat J_{n,m}k$ bounds the displacement to a target equilibrium,
and $\Lambda_{n,m}k$ bounds the diameter of the residual-equilibrium set.
We now derive these constants and choose $\mathfrak B_{n,m}$ accordingly.
Specializing Theorem~\ref{thm:mixed-rank-proximity} to a halving step, put
\begin{align}
 \theta_m&:=2(m^2+1)+\frac12,\label{eq:restore-theta}\\
 \widehat G_{n,m}
 &:=n\left[
 m\theta_m
 +\frac m2\bigl(H_m+(m^2+1)+2m\theta_m\bigr)\right.\notag\\
 &\hspace{36mm}\left.
 +\frac{m(m+1)}2\left(\frac32+\theta_m\right)
 \right],
 \label{eq:restore-Ghat}\\
 \widehat J_{n,m}
 &:=\theta_m
 +\frac m2\bigl(H_m+(m^2+1)+2m\theta_m\bigr)\notag\\
 &\quad+\frac{m+1}{2}
   \left((m-1)\left(\widehat G_{n,m}+\frac32+\theta_m\right)\right).
 \label{eq:restore-Jhat}
\end{align}
Indeed, for $h=k/2$, choose $\ell=h$.  By
\eqref{eq:successive-rank-distance}, $\Delta\le(m^2+1)k$, so
\[
 \eta=2\Delta+\ell\le\theta_mk.
\]
Substitution in \eqref{eq:mixed-rank-G}--\eqref{eq:mixed-rank-J} shows that
the mixed-rank theorem gives
\begin{equation}
\label{eq:restore-mixed-rank-bound}
 \max_{\substack{i\in N\\e\in E}}|x_{i,e}^k-x_{i,e}^{k/2}|
 \le\widehat J_{n,m}k
\end{equation}
for equilibria on consecutive effective ranks.

We also need the diameter of the equilibrium set of one compatible packet
game.  Define
\begin{equation}
\label{eq:restore-lambda-new}
 A_{n,m}:=\frac{nm(m+1)}2,
 \qquad
 \Lambda_{n,m}:=(m-1)(A_{n,m}+1).
\end{equation}
Corollary~\ref{cor:equilibrium-lattice-diameter} says that any two equilibria of the same
$h$-integral game differ in each player-resource coordinate by at most
$2\Lambda_{n,m}h$.  Finally choose the integer buffer
\begin{equation}
\label{eq:rank-restore-buffer}
 \mathfrak B_{n,m}:=\left\lceil
 \max\left\{m^2+1,
       \widehat J_{n,m}+\Lambda_{n,m}+\frac12\right\}
 \right\rceil
\end{equation}
and put $Q_{n,m}:=1+2m\mathfrak B_{n,m}$.

\begin{center}
\fbox{%
\begin{minipage}{0.93\linewidth}
\algorithmheading{alg:rank-restore}{$\mathsf{RankRestore}(x^k,k,h;f,g,c)$}

\smallskip
\noindent\textbf{Input:} Consecutive dyadic packet sizes $h=k/2$; the effective
ranks $f_i=\sigma_{i,k}$ and $g_i=\sigma_{i,h}$, given by value oracles;
affine cost data $c$; and a $k$-integral equilibrium
$x^k\in\prod_{i\in N}B(f_i)$.

\smallskip
\noindent\textbf{Output:} An $h$-integral equilibrium
$x^h\in\prod_{i\in N}B(g_i)$.

\begin{enumerate}[label=\arabic*.,leftmargin=2.2em,itemsep=2pt,topsep=4pt]
\item Freeze the coordinatewise lower vector
\begin{equation}
\label{eq:rank-restore-floor}
 w_{i,e}:=\max\{x_{i,e}^k-\mathfrak B_{n,m}k,0\}
 \qquad(i\in N,\ e\in E).
\end{equation}
\item Form the target residual ranks $g_i^{w_i}$ using
\eqref{eq:residual-rank-new} and the shifted costs using
\eqref{eq:shifted-cost-new}.
\item Compute an $h$-integral equilibrium $z^h$ of the residual game with
$\PAlg$.
\item Return
\[
 x^h:=w+z^h.
\]
\end{enumerate}
\end{minipage}}
\end{center}

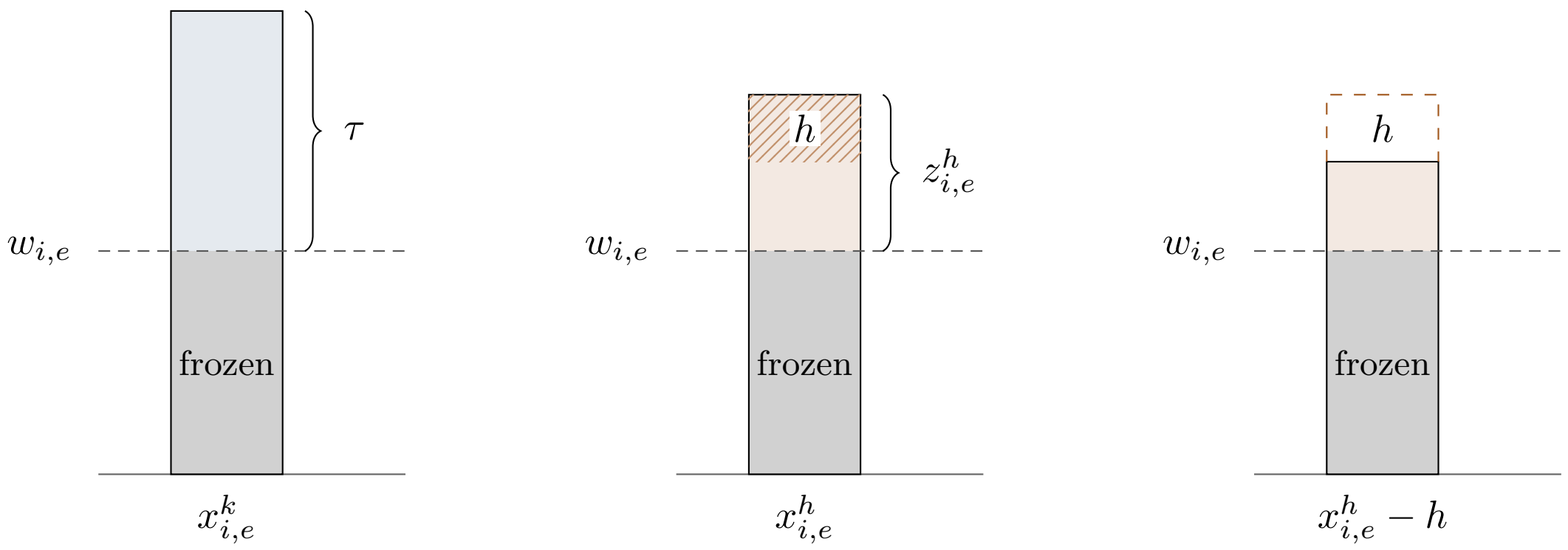
\begin{figure}[!htbp]
\centering
\begin{subfigure}[t]{.31\linewidth}
\centering\vspace{0pt}
\begin{tikzpicture}[x=1cm,y=1cm,font=\small]
  \path[use as bounding box] (-.65,-.65) rectangle (3.45,4.35);
  \draw[black!60] (.25,0)--(3,0);
  \fill[black!18] (.9,0) rectangle (1.9,2);
  \fill[proofblue!12] (.9,2) rectangle (1.9,4.15);
  \draw (.9,0) rectangle (1.9,4.15);
  \draw[densely dashed,black!65] (.25,2)--(3,2);
  \node[anchor=east] at (.15,2) {$w_{i,e}$};
  \node[font=\footnotesize] at (1.4,1) {frozen};
  \draw[decorate,decoration={brace,amplitude=4pt}] (2.1,4.15)--(2.1,2)
    node[midway,right=6pt] {$\tau$};
  \node at (1.4,-.4) {$x^k_{i,e}$};
\end{tikzpicture}
\caption{Before restore.}
\label{fig:buffer-before}
\end{subfigure}\hfill
\begin{subfigure}[t]{.31\linewidth}
\centering\vspace{0pt}
\begin{tikzpicture}[x=1cm,y=1cm,font=\small]
  \path[use as bounding box] (-.65,-.65) rectangle (3.45,4.35);
  \draw[black!60] (.25,0)--(3,0);
  \fill[black!18] (.9,0) rectangle (1.9,2);
  \fill[prooforange!13] (.9,2) rectangle (1.9,3.4);
  \draw (.9,0) rectangle (1.9,3.4);
  \fill[pattern=north east lines,pattern color=prooforange!65]
    (.9,2.8) rectangle (1.9,3.4);
  \draw[densely dashed,black!65] (.25,2)--(3,2);
  \node[anchor=east] at (.15,2) {$w_{i,e}$};
  \node[font=\footnotesize] at (1.4,1) {frozen};
  \draw[decorate,decoration={brace,amplitude=4pt}] (2.1,3.4)--(2.1,2)
    node[midway,right=6pt] {$z^h_{i,e}$};
  \node[fill=white,inner sep=1pt] at (1.4,3.1) {$h$};
  \node at (1.4,-.4) {$x^h_{i,e}$};
\end{tikzpicture}
\caption{After restore.}
\label{fig:buffer-after}
\end{subfigure}\hfill
\begin{subfigure}[t]{.31\linewidth}
\centering\vspace{0pt}
\begin{tikzpicture}[x=1cm,y=1cm,font=\small]
  \path[use as bounding box] (-.65,-.65) rectangle (3.45,4.35);
  \draw[black!60] (.25,0)--(3,0);
  \fill[black!18] (.9,0) rectangle (1.9,2);
  \fill[prooforange!13] (.9,2) rectangle (1.9,2.8);
  \draw (.9,0) rectangle (1.9,2.8);
  \draw[prooforange,dashed] (.9,2.8)--(.9,3.4)--(1.9,3.4)--(1.9,2.8);
  \draw[densely dashed,black!65] (.25,2)--(3,2);
  \node[anchor=east] at (.15,2) {$w_{i,e}$};
  \node[font=\footnotesize] at (1.4,1) {frozen};
  \node at (1.4,3.1) {$h$};
  \node at (1.4,-.4) {$x^h_{i,e}-h$};
\end{tikzpicture}
\caption{After removing an $h$-packet.}
\label{fig:buffer-removal}
\end{subfigure}
\par\medskip
{\small
\[
 \underbrace{\widehat J_{n,m}k}_{\text{equilibrium displacement}}
 +\underbrace{\Lambda_{n,m}k}_{\text{residual diameter}}
 +\underbrace{h}_{\text{one packet}}\le\tau.
\]
$z^h_{i,e}\ge\tau-(\widehat J_{n,m}+\Lambda_{n,m})k\ge h
\quad\Longrightarrow\quad x^h_{i,e}-h\ge w_{i,e}$.\par}
\caption{The source buffer in the proof of
Lemma~\ref{lem:rank-changing-restore}, for a coordinate with $w_{i,e}>0$.
The frozen amount is $w_{i,e}=x^k_{i,e}-\tau$, where
$\tau=\mathfrak B_{n,m}k$.  Proximity and the residual-equilibrium diameter
bound leave at least one $h$-packet in the residual allocation $z^h_{i,e}$.
After a feasible target exchange removes this packet, the source coordinate
remains at or above $w_{i,e}$.  For $w_{i,e}=0$, the same conclusion follows
from feasibility of the target exchange.}
\label{fig:restore-buffer}
\end{figure}

Figure~\ref{fig:restore-buffer} illustrates how the buffer keeps a removable
packet above the frozen allocation.

\begin{lemma}[Correctness of the restore procedure]
\label{lem:rank-changing-restore}
Let $k$ and $h=k/2$ be consecutive dyadic scales, let
$f_i=\sigma_{i,k}$ and $g_i=\sigma_{i,h}$, for every $i\in N$, be the
effective rounded ranks of Theorem~\ref{thm:compatible-rounding} applied
playerwise, and let $x^k$ be an equilibrium of the $k$-integral game on
the strategy spaces $(B(f_i))_{i\in N}$.  Then the output $x^h$ of
$\RankRestore(x^k,k,h;f,g,c)$ is an equilibrium of the target
$h$-integral game on the strategy spaces $(B(g_i))_{i\in N}$.
Each residual player has at most $Q_{n,m}$ packets.  The running time of
one restore call is polynomial in $n,m$, the binary lengths of the cost
coefficients, of the rank-oracle outputs, and of $k$, and the running
time of submodular-function minimization.  Numerical rank values enter only
through their binary length.
\end{lemma}

\begin{proof}
Write $\tau=\mathfrak B_{n,m}k$.  For every $i\in N$, put
\[
 \Delta_i:=\max_{S\subseteq E}|f_i(S)-g_i(S)|.
\]
By Theorem~\ref{thm:compatible-rounding}, $\tau\ge\Delta_i$ for every
$i\in N$.  For every
$i\in N$ and $S\subseteq E$,
\begin{align*}
 w_i(S)
 &=x_i^k(S)-\sum_{e\in S}\min\{x_{i,e}^k,\tau\}\\
 &\le x_i^k(S)-\min\{x_i^k(S),\tau\}\\
 &=\max\{x_i^k(S)-\tau,0\}\\
 &\le\max\{f_i(S)-\tau,0\}\le g_i(S).
\end{align*}
This proves $w_i\in P(g_i)$.  Its coordinates are multiples of $k$, and hence of
$h$, so Lemma~\ref{lem:residual-polymatroid-new} makes $g_i^{w_i}$ an
$h$-integral polymatroid rank.

Choose any equilibrium $\widehat x^h$ of the target game.  The bound
\eqref{eq:restore-mixed-rank-bound} and
$\tau\ge\widehat J_{n,m}k$ (by \eqref{eq:rank-restore-buffer}) imply
$\widehat x^h\ge w$ coordinatewise: this is
immediate where $w_{i,e}=0$, while otherwise
\[
 \widehat x_{i,e}^h
 \ge x_{i,e}^k-\widehat J_{n,m}k
 \ge x_{i,e}^k-\tau=w_{i,e}.
\]
The profile $\widehat z^h:=\widehat x^h-w$ is feasible in the residual game.
Every residual deviation is a target-game deviation, and
\eqref{eq:shifted-equivalence-new} shows that $\widehat z^h$ is a residual
equilibrium.

Let $z^h$ be the equilibrium returned by $\PAlg$.  The
residual ranks $g_i^{w_i}$ are $h$-compatible by
Lemma~\ref{lem:residual-polymatroid-new}, and the shifted costs remain affine
with the same positive slopes.  The residual game satisfies the hypotheses of
Corollary~\ref{cor:equilibrium-lattice-diameter}, whose diameter bound gives
\begin{equation}
\label{eq:rank-restore-residual-distance}
 |z_{i,e}^h-\widehat z_{i,e}^h|
 \le2\Lambda_{n,m}h=\Lambda_{n,m}k.
\end{equation}
Whenever $w_{i,e}>0$,
\[
 \widehat z_{i,e}^h
 =\widehat x_{i,e}^h-w_{i,e}
 \ge \tau-\widehat J_{n,m}k,
\]
and therefore, by \eqref{eq:rank-restore-buffer} and
\eqref{eq:rank-restore-residual-distance},
\begin{equation}
\label{eq:rank-restore-source-buffer}
 z_{i,e}^h
 \ge \tau-(\widehat J_{n,m}+\Lambda_{n,m})k
 \ge k/2=h.
\end{equation}

Consider any feasible target one-packet move
$x_i^h+h(\mathbf1_s-\mathbf1_e)\in B(g_i)$.  If $w_{i,e}=0$, then
$z_{i,e}^h=x_{i,e}^h\ge h$; if $w_{i,e}>0$, the same conclusion follows
from \eqref{eq:rank-restore-source-buffer}.  Hence the residual vector after
the move is nonnegative.  Lemma~\ref{lem:residual-polymatroid-new} then
shows that it is a feasible packet move in $B(g_i^{w_i})$.  The shifted discrete
marginal costs equal the original target marginal costs at $w+z^h$.
Residual one-packet optimality therefore implies target one-packet
optimality for every feasible packet move.  By compatibility, the criterion
\eqref{eq:packet-optimality} shows that $x^h$ is a target equilibrium.

It remains to bound the residual demand.  Consecutive rounded base values
satisfy
\[
 0\le g_i(E)-f_i(E)=d_{i,h}-d_{i,k}\le h.
\]
Since $x_i^k(E)=f_i(E)$ and
$x_{i,e}^k-w_{i,e}\le \tau$,
\begin{align*}
 g_i^{w_i}(E)
 &=g_i(E)-w_i(E)\\
 &=g_i(E)-f_i(E)+x_i^k(E)-w_i(E)\\
 &\le h+m\tau.
\end{align*}
After division by $h=k/2$, this is at most
$1+2m\mathfrak B_{n,m}=Q_{n,m}$ packets.  The implementation above and the
$O(n^2mQ_{n,m}^3)$ restricted best-response bound of
Proposition~\ref{prop:hkp-algorithm} show that one restore call uses
$O(n^3m^2Q_{n,m}^3)$ submodular-minimization calls and polynomially many exact
arithmetic operations.  Throughout a call, all loads are dyadic rationals of
absolute value at most $nR_{\max}$, and their denominators divide the current packet
denominator.  Their bit lengths, the shifted intercepts, and every value
passed to submodular minimization have length polynomial in the input bit
length and the scale index.
\end{proof}

\FloatBarrier

\subsection{The complete packet-halving algorithm}

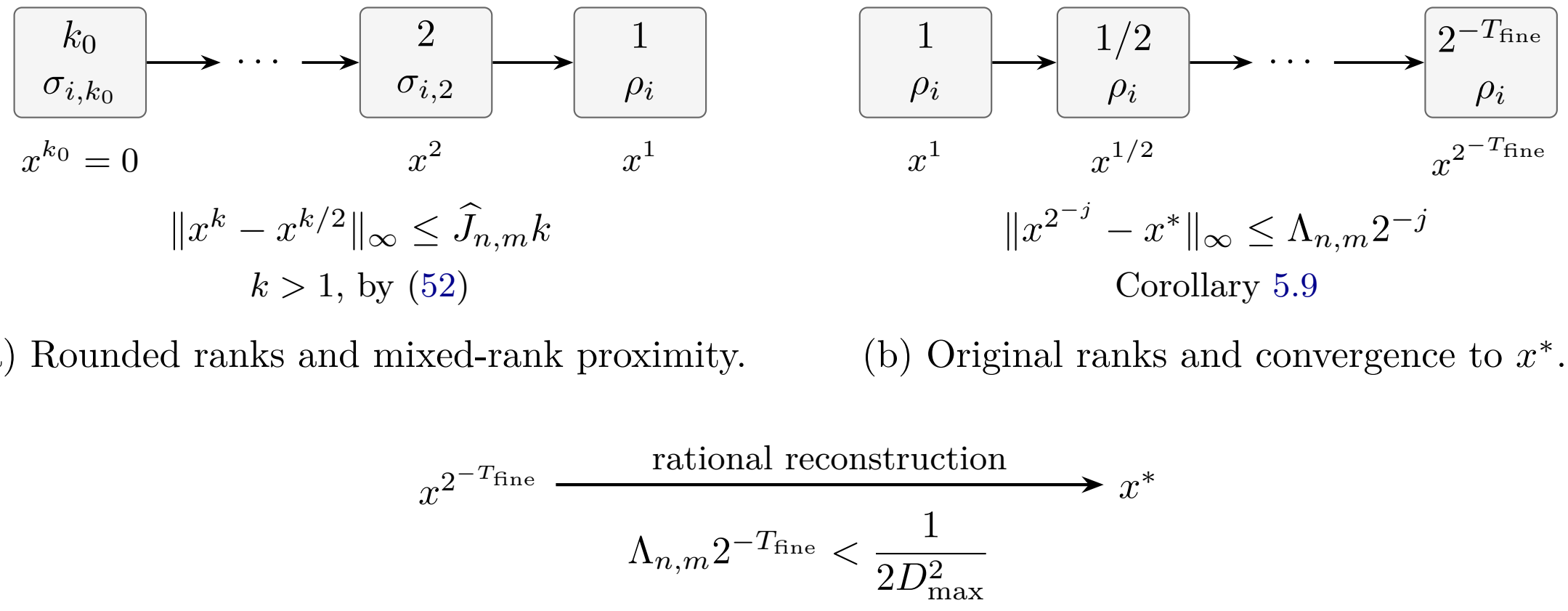
\begin{figure}[!htbp]
\centering
\begin{subfigure}[t]{.48\linewidth}
\centering\vspace{0pt}
\begin{tikzpicture}[
  phase/.style={draw=black!60,rounded corners=2pt,fill=black!4,
    align=center,font=\small,inner sep=3pt,minimum width=12mm,
    minimum height=10mm},
  scaling arrow/.style={-{Stealth[length=1.8mm]},line width=.7pt},
  font=\small]
  \path[use as bounding box] (0,-2.3) rectangle (6.8,.75);
  \node[phase] (start) at (.85,0) {$k_0$\\$\sigma_{i,k_0}$};
  \node (dots) at (2.5,0) {$\cdots$};
  \node[phase] (two) at (4,0) {$2$\\$\sigma_{i,2}$};
  \node[phase] (one) at (5.95,0) {$1$\\$\rho_i$};
  \draw[scaling arrow] (start)--(dots);
  \draw[scaling arrow] (dots)--(two);
  \draw[scaling arrow] (two)--(one);
  \node[font=\footnotesize] at (.85,-.85) {$x^{k_0}=0$};
  \node[font=\footnotesize] at (4,-.85) {$x^2$};
  \node[font=\footnotesize] at (5.95,-.85) {$x^1$};
  \node at (3.4,-1.5)
    {$\|x^k-x^{k/2}\|_\infty\le\widehat J_{n,m}k$};
  \node[font=\footnotesize] at (3.4,-2.05)
    {$k>1$, by \eqref{eq:restore-mixed-rank-bound}};
\end{tikzpicture}
\caption{Rounded ranks and mixed-rank proximity.}
\label{fig:scaling-coarse}
\end{subfigure}\hfill
\begin{subfigure}[t]{.48\linewidth}
\centering\vspace{0pt}
\begin{tikzpicture}[
  phase/.style={draw=black!60,rounded corners=2pt,fill=black!4,
    align=center,font=\small,inner sep=3pt,minimum width=12mm,
    minimum height=10mm},
  scaling arrow/.style={-{Stealth[length=1.8mm]},line width=.7pt},
  font=\small]
  \path[use as bounding box] (0,-2.3) rectangle (6.8,.75);
  \node[phase] (one) at (.75,0) {$1$\\$\rho_i$};
  \node[phase] (half) at (2.55,0) {$1/2$\\$\rho_i$};
  \node (dots) at (4.1,0) {$\cdots$};
  \node[phase] (last) at (5.9,0) {$2^{-T_{\rm fine}}$\\$\rho_i$};
  \draw[scaling arrow] (one)--(half);
  \draw[scaling arrow] (half)--(dots);
  \draw[scaling arrow] (dots)--(last);
  \node[font=\footnotesize] at (.75,-.85) {$x^1$};
  \node[font=\footnotesize] at (2.55,-.85) {$x^{1/2}$};
  \node[font=\footnotesize] at (5.9,-.85) {$x^{2^{-T_{\rm fine}}}$};
  \node at (3.4,-1.5)
    {$\|x^{2^{-j}}-x^*\|_\infty\le\Lambda_{n,m}2^{-j}$};
  \node[font=\footnotesize] at (3.4,-2.05)
    {Corollary~\ref{cor:continuous-proximity}};
\end{tikzpicture}
\caption{Original ranks and convergence to $x^*$.}
\label{fig:scaling-fine}
\end{subfigure}
\par\bigskip
\begin{subfigure}[t]{\linewidth}
\centering
\begin{tikzpicture}[font=\small]
  \node (packet) at (0,0) {$x^{2^{-T_{\rm fine}}}$};
  \node (exact) at (6,0) {$x^*$};
  \draw[-{Stealth[length=2mm]},line width=.8pt]
    (packet)--node[above,font=\footnotesize] {rational reconstruction} (exact);
  \node at (3,-.65)
    {$\Lambda_{n,m}2^{-T_{\rm fine}}<\dfrac{1}{2D_{\max}^2}$};
\end{tikzpicture}
\caption{Exact recovery at the depth specified in \eqref{eq:recovery-depth-new}.}
\label{fig:scaling-recovery}
\end{subfigure}
\caption{From coarse packets to the continuous equilibrium.
Each box lists a packet size and player $i$'s effective rank; the profile
below it is an equilibrium at that scale.  Each halving in
(\subref{fig:scaling-coarse}) and (\subref{fig:scaling-fine}) uses
$\RankRestore$ and solves a residual game with at most $Q_{n,m}$
packets per player.  Mixed-rank proximity controls the changing ranks;
from unit scale on, $\sigma_{i,k}=\rho_i$.
The denominator bound of Lemma~\ref{lem:rational-size-new} permits the
final reconstruction in (\subref{fig:scaling-recovery}).}
\label{fig:scaling-chain}
\end{figure}

The following procedure is the polymatroid counterpart of the packet-halving
algorithm of Harks and Timmermans \cite{HarksT22}, with the rounding transition and
$\RankRestore$ replacing its separate restore-and-add phases.
Figure~\ref{fig:scaling-chain} follows the rounded ranks to unit scale,
the subsequent refinement with the original ranks, and exact recovery of
the continuous equilibrium.

Let the original ranks $\rho_i$ be integral and value-oracle accessible, and
recall $R_{\max}=\max_{i\in N}\rho_i(E)$.  Choose
\begin{equation}
\label{eq:initial-packet}
 k_0:=2^{\lceil\log_2(R_{\max}+1)\rceil}.
\end{equation}
Then $k_0>R_{\max}$, so every effective rank $\sigma_{i,k_0}$ has base value zero
and the empty profile is its unique equilibrium.  Starting from this profile,
apply Lemma~\ref{lem:rank-changing-restore} along
\[
 k_0,\frac{k_0}{2},\frac{k_0}{4},\ldots,1.
\]
At unit scale Theorem~\ref{thm:compatible-rounding} gives
$\sigma_{i,1}=\rho_i$.  The algorithm reaches an equilibrium of the
original unit-packet game through residual games with at most $Q_{n,m}$
packets per player.  Continuing along
\[
 1,\frac12,\frac14,\ldots,2^{-T_{\rm fine}}
\]
uses the original ranks at every phase and produces an equilibrium of the
original $2^{-T_{\rm fine}}$-integral game.

Each restore call incorporates both the finer packet size and the new base
value.  As proved in Lemma~\ref{lem:rank-changing-restore}, the additional
demand is already included in the residual packet bound $Q_{n,m}$.

\begin{center}
\fbox{%
\begin{minipage}{0.93\linewidth}
\algorithmheading{alg:rounding-halving}{\textsc{Rounding-Halving}$(\rho,c,T_{\rm fine})$}

\begin{enumerate}[label=\arabic*.,leftmargin=2.2em,itemsep=2pt,topsep=4pt]
 \item Compute $k_0$ from \eqref{eq:initial-packet}; set $x=0$ and $k=k_0$.
 \item While $k>1$, construct the target rank $\sigma_{k/2}$ by
       \eqref{eq:effective-rounding}, replace $x$ by
       $\RankRestore(x,k,k/2;\sigma_k,\sigma_{k/2},c)$, and set
       $k\leftarrow k/2$.
 \item Perform $T_{\rm fine}$ further restore steps with rank $\rho$ at both scales.
 \item Return the resulting $2^{-T_{\rm fine}}$-integral equilibrium.
\end{enumerate}
\end{minipage}}
\end{center}

Put
\[
 P_{\rm phases}:=\log_2 k_0+T_{\rm fine}=O(\log(R_{\max}+1)+T_{\rm fine})
\]
for the number of phases.  Every phase has at most $Q_{n,m}$ residual packets
per player.  The number of restricted best-response updates is therefore
\[
 O\bigl(P_{\rm phases}n^2mQ_{n,m}^3\bigr),
\]
and the implementation above uses
$O(P_{\rm phases}n^3m^2Q_{n,m}^3)$ submodular-minimization calls.  If $L_{\rm max}$ is the
maximum bit length of an oracle value or exact rational number occurring up
to scale $2^{-T_{\rm fine}}$, then
\[
 L_{\rm max}=\operatorname{poly}\bigl(
 L_{\rm in}+\log(R_{\max}+1)+T_{\rm fine}+\log(n+m)\bigr),
\]
where $L_{\rm in}$ is the maximum binary length of an input cost coefficient
or rank-oracle output.  An overall running-time bound is
\begin{equation}
\label{eq:complete-halving-runtime}
 O\left(
 P_{\rm phases}n^3m^2Q_{n,m}^3\,\mathsf{SFM}(m,L_{\rm max})
 +\operatorname{poly}(P_{\rm phases},n,m,Q_{n,m},L_{\rm max})
 \right).
\end{equation}
This is polynomial in the stated oracle model; the numerical rank $R_{\max}$ appears
only through its binary length $\log(R_{\max}+1)$.

To compute an equilibrium for an arbitrary compatible rational packet size,
first rescale that size to one and then stop the chain at unit scale.

\begin{theorem}[Exact $k$-integral equilibria]
\label{thm:k-integral-computation}
Let the ranks $\rho_i$ be integral and value-oracle accessible, let the costs
be player-specific rational affine with positive slopes, and let $k>0$ be
rational and compatible with every $\rho_i$.  Then a $k$-integral equilibrium
can be computed in time polynomial in $n$, $m$, the binary length of the
input, and the binary length of $k$, in the same oracle model as
Theorem~\ref{thm:exact-computation-new}.  In particular the running time is
polynomial even though a player may own $\rho_i(E)/k$ packets, a number that
is in general exponential in the input length.
\end{theorem}

\begin{proof}
Rescale the instance by $1/k$.  Put $\rho_i^{(k)}:=\rho_i/k$; by
compatibility this is again an integral polymatroid rank function, one query
to the original oracle produces one of its values, and its binary length
exceeds that of $\rho_i$ by at most the binary length of $k$.  Put
\[
 c^{(k)}_{i,e}(z):=c_{i,e}(kz)=(ka_{i,e})z+b_{i,e},
\]
which is again rational affine with positive slope.  The map $x\mapsto x/k$ is
a bijection from the $k$-integral profiles of the original game onto the
integral profiles of the rescaled game, and
\[
 \sum_{e\in E}x_{i,e}c_{i,e}(x_e)
 =k\sum_{e\in E}z_{i,e}c^{(k)}_{i,e}(z_e),
 \qquad z=x/k,
\]
so every player's cost is multiplied by the same positive constant $k$ and the
two games have the same set of best responses.  Hence $x$ is a $k$-integral
equilibrium of the original game if and only if $x/k$ is an integral
equilibrium of the rescaled game.

Now run \RoundingHalving{} on the rescaled instance with $T_{\rm fine}=0$, that is,
stop at unit scale.  By Theorem~\ref{thm:compatible-rounding} the effective
rank at unit scale is $\rho^{(k)}$ itself, so the last phase returns an
integral equilibrium of the rescaled game, and multiplying it by $k$ gives a
$k$-integral equilibrium of the original one.  The number of phases is
$O(\log(R_{\max}/k+1))$, which is polynomial in the binary lengths of $R_{\max}$ and $k$,
and every phase obeys the residual packet bound $Q_{n,m}$ of
Lemma~\ref{lem:rank-changing-restore}.  The running-time estimate
\eqref{eq:complete-halving-runtime} applies verbatim with $T_{\rm fine}=0$ and $R_{\max}$
replaced by $R_{\max}/k$.
\end{proof}

\begin{remark}[Size of the constants]
\label{rem:constant-size}
The polynomial degree comes from the rounding error, the mixed-rank
path count, and the residual solver.  The correction $\Psi_m$ gives
rank distance $O(m^2k)$ at a halving step.  Substituting this into the
individual bound $O(nm^5(k+h)+nm^4\Delta)$ gives $O(nm^6k)$;
here the mixed-rank proof uses the path count to accommodate missing
reverse edges.  From \eqref{eq:restore-theta}--\eqref{eq:restore-Jhat} one gets
$\theta_m=\Theta(m^2)$, $\widehat G_{n,m}=O(nm^4)$ and
$\widehat J_{n,m}=O(nm^6)$, hence $\mathfrak B_{n,m}=O(nm^6)$ and
$Q_{n,m}=O(nm^7)$.  The residual packet count therefore enters
\eqref{eq:complete-halving-runtime} as $Q_{n,m}^3=O(n^3m^{21})$.
Multiplying the $O(n^2mQ_{n,m}^3)$ restricted best responses by the $O(nm)$
submodular minimizations per update and by the number of phases yields
$O(P_{\rm phases}n^6m^{23})$ submodular minimizations.
Determining sharper dependence on $n$ and $m$ is a further quantitative
question.
\end{remark}

Exact recovery uses a polynomial bound on the binary length of the
continuous equilibrium coordinates.
An explicit value of $H$ below is obtained by multiplying the affine
marginal-cost map by the product of the denominators of its rational
coefficients and then taking the maximum of their resulting absolute integer
values, that product, $R_{\max}$, and $2$.  Monotonicity gives
$0\le\rho_i(S)\le\rho_i(E)\le R_{\max}$, so this $H$ bounds every rank value and is
computable from the cost input and the $n$ oracle queries at $E$.  Its binary
length is polynomial in the input length.

\begin{lemma}[Rational size of the continuous equilibrium]
\label{lem:rational-size-new}
Let $d=nm$.  After clearing denominators, suppose that the integer
coefficients of the affine marginal-cost map and all rank values have
absolute value at most $H\ge2$.  Every coordinate of the unique continuous
equilibrium $x^*$ is rational with denominator at most
\begin{equation}
\label{eq:denominator-bound-new}
 D_{\max}:=(4dH)^{4d}.
\end{equation}
\end{lemma}

\begin{proof}
Write the rational strategy polytope in the form
\[
 X=\{x\in\mathbb R^d:Ax\le r,\ Ex=s\}.
\]
The rows of $A$ are the subset-constraint normals and the nonnegativity
normals, while $Ex=s$ contains the $n$ player-wise base equalities.  Thus the
entries of $A,E$ belong to $\{0,\pm1\}$ and the entries of $r,s$ are rank
values.  After clearing denominators and rescaling normal-cone multipliers,
the marginal-cost map has the form $\mu(x)=Mx+b$ with integral $M,b$, and all
entries of $A,E,M,b,r,s$ have absolute value at most $H$.

The equilibrium condition is the affine variational inequality
\[
 0\in Mx+b+N_X(x),
 \qquad X=\prod_iB(\rho_i).
\]
Let
\[
 I:=\{j:A_jx^*=r_j\}
\]
be the active inequality set.  The normal-cone formula gives multipliers
$\lambda\ge0$ and a free multiplier $\nu$ such that
\[
 Mx^*+b+A_I^\mathsf T\lambda+E^\mathsf T\nu=0.
\]
By conic Carath\'eodory, the conic vector
$A_I^\mathsf T\lambda$ has a representation using a set
$J\subseteq I$ of at most $d$ rows.  Retain only these rows and split the free
multiplier as $\nu=\nu^+-\nu^-$ with $\nu^+,\nu^-\ge0$.  Consider the rational
polyhedron
\[
\begin{aligned}
 \mathcal Q_J:=\bigl\{(x,\lambda_J,\nu^+,\nu^-):\;&
 Ax\le r,\quad Ex=s,\quad A_Jx=r_J,\\
 &Mx+b+A_J^\mathsf T\lambda_J
          +E^\mathsf T(\nu^+-\nu^-)=0,\\
 &\lambda_J,\nu^+,\nu^-\ge0\bigr\}.
\end{aligned}
\]
It is nonempty because it contains the representation just constructed for
$x^*$.  Conversely, at every point of $\mathcal Q_J$, the rows in $J$ are
active and
\[
 -Mx-b=A_J^\mathsf T\lambda_J+E^\mathsf T(\nu^+-\nu^-)
       \in N_X(x).
\]
Hence its $x$-component solves the variational inequality and, by uniqueness,
equals $x^*$.  In particular, the $x$-projection of $\mathcal Q_J$ is the
singleton $\{x^*\}$.

The polyhedron $\mathcal Q_J$ contains no affine line: along such a line
the $x$-component would have to remain fixed, and a nonzero direction in
any multiplier coordinate would violate nonnegativity in one direction.
Being nonempty and containing no line, it has a vertex.

There are $N=d+|J|+2n\le4d$ variables.  At a vertex, choose $N$ linearly
independent active constraints and let $B$ be their coefficient matrix.
This is a nonsingular integer matrix with entries bounded in absolute value
by $H$, and its right-hand side is integral.  Cramer's rule therefore makes
each coordinate denominator a divisor of $|\det B|$.  Each row of $B$ has
Euclidean norm at most $\sqrt N H$, so Hadamard's inequality gives
\[
 |\det B|\le(\sqrt N H)^N\le(4dH)^{4d}=D_{\max}.
\]
The $x$-coordinates of this vertex equal $x^*$, proving the claimed bound.
\end{proof}

By Corollary~\ref{cor:continuous-proximity}, an equilibrium $x^{2^{-T_{\rm fine}}}$
of the exact-rank packet game satisfies
\[
 \|x^{2^{-T_{\rm fine}}}-x^*\|_\infty
 \le\Lambda_{n,m}2^{-T_{\rm fine}}.
\]
Distinct rationals with denominators at most $D_{\max}$ differ by at least $D_{\max}^{-2}$.
Choose
\begin{equation}
\label{eq:recovery-depth-new}
 T_{\rm fine}>\log_2(2\Lambda_{n,m}D_{\max}^2).
\end{equation}
Continued-fraction reconstruction then recovers every coordinate of $x^*$ as
the unique rational of denominator at most $D_{\max}$ in the corresponding error
interval.  To verify the reconstructed profile exactly, first check
$x_i\ge0$ and $x_i(E)=\rho_i(E)$ for every player, and then minimize
$\rho_i(S)-x_i(S)$ over $S\subseteq E$ to check all subset constraints.
Finally, minimize the linear objective $\langle\mu_i(x),z_i\rangle$ over
$B(\rho_i)$ by the polymatroid greedy algorithm and compare its minimum
with $\langle\mu_i(x),x_i\rangle$ to verify the best-response variational
inequality.

\begin{theorem}[Exact equilibrium computation for arbitrary integral ranks]
\label{thm:exact-computation-new}
Let the player-specific integral polymatroid ranks be given by value oracles,
with binary-encoded values, and let the player-specific affine costs be
rational with positive slopes.  Then the unique continuous Nash equilibrium
can be computed exactly in time polynomial in $n,m$, the binary lengths of
the cost coefficients and rank-oracle outputs, and the running time of
submodular-function minimization.  More generally, an equilibrium of any
prescribed dyadic packet size $2^{-T_{\rm fine}}$ can be computed in the time
\eqref{eq:complete-halving-runtime}.  
\end{theorem}

\begin{proof}
The empty initialization is valid because $k_0>R_{\max}$.  Induction using
Lemma~\ref{lem:rank-changing-restore} proves that every coarse phase returns
an equilibrium for the next effective rounded rank.  The rounded rank equals the
original rank at unit scale and below, so the remaining phases return exact
packet equilibria of the original game.  The phase count and residual packet
bound give \eqref{eq:complete-halving-runtime}.  Both $\log(R_{\max}+1)$ and $\log D_{\max}$, and hence the required depth $T_{\rm fine}$ in
\eqref{eq:recovery-depth-new}, are polynomial in the binary input length.  Rational reconstruction and exact verification are
polynomial.  The reconstructed profile is the continuous equilibrium, which
is unique by Theorem~\ref{thm:uniqueness}.
\end{proof}

\section{Open Problems}
\label{sec:open-problems}

The results suggest two directions for further study: allowing more general
interactions between players on a resource, and replacing separable convex
best-response objectives by discrete convex functions.  In both directions,
the main challenge is to extend the equilibrium proximity and scaling
arguments beyond the exchange properties of individual strategy spaces.

\subsection{Weighted games and multivariate congestion}

Takazawa \cite{Takazawa19} generalizes weighted matroid congestion games by
allowing resource costs to depend on the set of users, and proves existence
of pure Nash equilibria when these costs are nondecreasing under inclusion.
This motivates extensions of our proximity and scaling results.

For continuous allocations, constant player weights $w_i>0$ can be absorbed
by rescaling.  Consider
\[
 C_i^w(x)=\sum_{e\in E}x_{i,e}
 c_{i,e}\!\left(\sum_{j\in N}w_jx_{j,e}\right),
 \qquad x_i\in B(\rho_i).
\]
Setting $y_i=w_ix_i$ gives $y_i\in B(w_i\rho_i)$ and transforms
$w_iC_i^w$ into our original cost model.  Since positive rescaling of an
objective preserves best responses, Theorem~\ref{thm:uniqueness} already
covers this weighted continuous model under the same latency assumptions.

For affine latencies with positive slopes, rational weights, and rational
packet sizes $k,h$ compatible with the original ranks, this substitution
produces player-dependent packet sizes $w_i k,w_i h$.
The proof of Theorem~\ref{thm:mixed-proximity} extends to these sizes:
each exchange graph concerns one player, and the common threshold
$t=w_{\max}(k+h)$, where $w_{\max}:=\max_i w_i$, dominates every player's
$w_i(k+h)$, so the cycle inequalities remain valid.  On a common lattice
with unit $\ell$, Lemma~\ref{lem:bad-count}(i) gives $|Z|\le mt/\ell$;
the same aggregate and individual counting arguments then give
$\max_e|y_{i,e}^k-y_{i,e}^h|\le\Lambda_{n,m}t$ for any two existing equilibria.
Dividing by $w_i$ converts this bound to the original allocation units:
\[
 \max_{e\in E}|x_{i,e}^k-x_{i,e}^h|
 \le\Lambda_{n,m}\frac{w_{\max}}{w_i}(k+h),
 \qquad i\in N,
\]
where $\Lambda_{n,m}$ is defined in \eqref{eq:restore-lambda-new}.
\emph{For classes admitting pure equilibria, can weighted packet games be
solved in time polynomial in the binary input length, including that of the
weights?}  The bound above can leave a residual packet count proportional
to a weight ratio, which may be exponential in that length.  Sharper bounds
and algorithms that handle unequal physical packet sizes are the remaining
questions.

A further extension, motivated by Takazawa \cite{Takazawa19}, allows
multivariate resource costs,
\[
 C_i(x)=\sum_{e\in E}F_{i,e}(x_{1,e},\ldots,x_{n,e}).
\]
An aggregate-load comparison then need not order marginal costs.
\emph{Which restrictions on interactions between coordinates preserve our
uniqueness and proximity arguments?}  Convexity in a player's own coordinate
must be complemented by conditions permitting comparisons across opponents'
profiles.

\subsection{Equilibrium proximity in discrete convex analysis}

Murota's discrete convex analysis \cite{Murota:2003} provides a natural
framework for the second direction.  At unit packet size, extending a
player's separable convex best-response objective by $+\infty$ outside its
integral polymatroid base gives an $\mathrm{M}$-convex function.
Allowing nonseparable $\mathrm{M}$-convex objectives, or
$\mathrm{M}^{\natural}$-convex objectives when total allocation may vary,
would extend the class of individual optimization problems.
The issue is how these objectives depend on the other players' strategies:
discrete convexity of each best-response problem alone does not guarantee
existence or uniqueness of a pure equilibrium.

There are relevant exchange and proximity results to build on.  Murota
\cite{Murota18Exchange} proves a multiple exchange property for valuated
matroids and $\mathrm{M}^{\natural}$-concave set functions; reversing signs
gives the corresponding cost convention.  Murota and Tamura
\cite{MurotaTamura04} establish proximity theorems for discrete convex
optimization.  In games, comparing two equilibria also changes the
opponents' strategies and hence the objective faced by each player.

The resulting question is: \emph{under what conditions on this dependence can
one obtain equilibrium proximity and a polynomial scaling algorithm for games
with $\mathrm{M}$-convex or $\mathrm{M}^{\natural}$-convex best-response
objectives?}  A possible route is to combine inequalities for function values
under multiple exchange with the support restrictions and quantitative
exchange bounds of Corollary~\ref{cor:support-restricted-strong-exchange}.
Such an argument must compare exchanges under different opponents' profiles.
It should also distinguish proximity to a suitably selected equilibrium from
a bound for every pair of equilibria: flat costs can already produce large
equilibrium sets.  For models with a specified continuous extension, a
further question is which strictness assumptions yield a unique continuous
limit and a quantitative rate of convergence as packet sizes vanish.

\section*{Acknowledgements}
The author started this work already several years ago and significant progress was achieved after Oki and Schwarcz~\cite{OkiSchwarcz25} derived their multi-exchange theorem for matroid bases. The resulting proof strategy, in particular,  the newly introduced compatible submodular rounding approach was 
developed by the author.
Generative AI (OpenAI ChatGPT/Codex) was used to assist with mathematical
review, exposition, LaTeX editing, and preparation of this manuscript.

\bibliographystyle{abbrv}
\bibliography{master-bib}

\end{document}